%% file: paper_SemComRL.tex
\documentclass[conference,comsoc]{IEEEtran}
\IEEEoverridecommandlockouts
\usepackage{cite}
\usepackage{amsmath,amssymb,amsfonts}
\usepackage{algorithmic}
\usepackage{graphicx}
\usepackage{textcomp}
\usepackage{xcolor}
\def\BibTeX{{\rm B\kern-.05em{\sc i\kern-.025em b}\kern-.08em
    T\kern-.1667em\lower.7ex\hbox{E}\kern-.125emX}}

\usepackage{bm}
\usepackage{tikz}
\usepackage{pgfplots}
\pgfplotsset{compat=newest}
\usetikzlibrary{calc, shapes.geometric, shapes.misc, arrows, matrix, fit, decorations.markings}
\usepackage[absolute, overlay]{textpos}

\input{definitions}
\input{plotopt}
\providecommand{\main}{.}

\begin{document}

\title{Model-free Reinforcement Learning of Semantic Communication by Stochastic Policy Gradient
	\thanks{This work was partly funded by the Federal State of Bremen and the University of Bremen as part of the Humans on Mars Initiative, by the German Ministry of Education and Research (BMBF) under grant 16KISK016 (Open6GHub), and by the German Research Foundation (DFG) under grant 500260669 (SCIL).}%
}

\author{\IEEEauthorblockN{Edgar Beck, Carsten Bockelmann and Armin Dekorsy}
	\IEEEauthorblockA{\textit{Department of Communications Engineering}\\
		\textit{University of Bremen}, Bremen, Germany \\
		Email: \{beck, bockelmann, dekorsy\}@ant.uni-bremen.de}
}

\maketitle

\begin{abstract}
	Following the recent success of Machine Learning tools in wireless communications, the idea of semantic communication by Weaver from 1949 has gained attention. It breaks with Shannon's classic design paradigm by aiming to transmit the meaning, i.e., semantics, of a message instead of its exact version, allowing for information rate savings.
	In this work, we apply the Stochastic Policy Gradient (SPG) to design a semantic communication system by reinforcement learning, separating transmitter and receiver, and not requiring a known or differentiable channel model -- a crucial step towards deployment in practice. Further, we motivate the use of SPG for both classic and semantic communication from the maximization of the mutual information between received and target variables. Numerical results show that our approach achieves comparable performance to a model-aware approach based on the reparametrization trick, albeit with a decreased convergence rate.
\end{abstract}

\begin{IEEEkeywords}
	Semantic communication, wireless networks, infomax, information bottleneck, machine learning, reinforcement learning, stochastic policy gradient, task-oriented.
\end{IEEEkeywords}

\section{Introduction}

To meet the unprecedented needs of 6G communication efficiency in terms of data rate, latency, and power, attention has been drawn to semantic communication~\cite{shannon_mathematical_1949, calvanese_strinati_6g_2021, gunduz_beyond_2023, beck_semantic_2023}. It aims to transmit the meaning of a message rather than its exact version, which has been the main focus of digital error-free system design so far~\cite{shannon_mathematical_1949}. Bao, Basu et al.~\cite{bao_towards_2011} were the first to define semantic information sources and channels to tackle the semantic design by conventional approaches arguing for the generality of Shannon's theory not only for the technical level but for semantic level design as Weaver~\cite{shannon_mathematical_1949}.

Recently, inspired by~\cite{shannon_mathematical_1949, bao_towards_2011} and the rise of Machine Learning (ML) in communications research, transformer-based Deep Neural Networks (DNNs), have been introduced to Auto Encoders (AEs) for text transmission to learn compressed hidden representations of semantic content, aiming to improve communication efficiency~\cite{xie_deep_2021}. In~\cite{lu_reinforcement_2022}, the authors suggest using semantic similarity as the objective function: As most semantic metrics are non-differentiable, they propose a self-critic Reinforcement Learning (RL) solution. Both~\cite{xie_deep_2021, lu_reinforcement_2022} improve performance especially at low SNR compared to classical digital transmissions with~\cite{lu_reinforcement_2022} being slightly superior.

This paper builds on our idea from~\cite{beck_semantic_2023}: There, we define semantic communication as the data-reduced, reliable transmission of semantic sources and cast its design as an Information Bottleneck (IB) problem extending~\cite{bao_towards_2011}. %
We apply our ML-based design Semantic INFOrmation TraNsmission and RecoverY (\sinfoni) to a distributed multipoint scenario, communicating meaning from multiple image sources to a single receiver for semantic recovery. Numerical results show that \sinfoni\ outperforms classical communication systems.

Semantic communication is a developing field: %
For a more in-depth survey, we refer the reader to, e.g.,~\cite{calvanese_strinati_6g_2021, gunduz_beyond_2023,beck_semantic_2023}. It remains still unclear how the approaches proposed so far can be implemented in practice which motivates the main contributions of this article:
\begin{itemize}
	\item We apply the Stochastic Policy Gradient (SPG) to train a semantic communication system, i.e., \sinfonirl, by RL. By this means, we separate transmitter and receiver, and do not require a known or differentiable channel model -- a crucial step towards deployment in practice.
	\item Further, we derive the application of the SPG for both classic and semantic communication from maximization of the mutual information between target and received variables compared to~\cite{aoudia_model-free_2019}.
	\item In particular, we investigate a task-oriented system model and a distributed application scenario with multiple sources and transmitters. By this means, our work distinguishes from the RL-based approach in~\cite{lu_reinforcement_2022} that was extended to handle non-differentiable channels at the time of writing.
	\item Further, the authors of~\cite{lu_reinforcement_2022} observed that training does not converge within their time limit to comparable results as the baseline approach in their setup for text transmission. We confirm the problem of slow convergence hinting at solution approaches and demonstrate feasibility in our scenario.
\end{itemize}
In the following, we revisit our theoretical framework from~\cite{beck_semantic_2023} in Sec.~\ref{sec:2}. For RL-based optimization, we introduce the SPG in Sec.~\ref{sec:3}. Finally, in Sec.~\ref{sec:4} and~\ref{sec:conclusion}, we provide one numerical example for \sinfoni\ application from~\cite{beck_semantic_2023} and summarize the main results, respectively.

\section{Semantic Communication Framework}
\label{sec:2}

\subsection{Semantic System Model}
\label{sec:23}

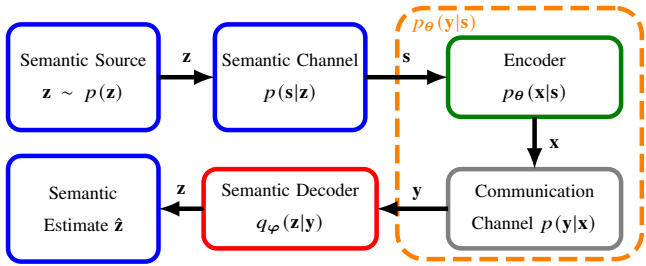
\begin{figure}[!t]%
	\centerline{\input{TikZ/com_system_model_basic_small_RL_v2.tikz}}
	\caption{Block diagram of the considered semantic system model.}
	\label{fig1:com_system_small}
\end{figure}

\subsubsection{Semantic Source and Channel}

First, we define our information-theoretic system model of semantic communication shown in Fig.~\ref{fig1:com_system_small}. Motivated by the approach of Bao, Basu et al.~\cite{bao_towards_2011}, we adopt the terminus of a semantic source as in~\cite{beck_semantic_2023} and describe it as a hidden target multivariate Random Variable (\rv) $\relvec\in\relset^{\Nrelvec\times 1}$ from domain $\relset$ of dimension $\Nrelvec$ distributed according to a probability density or mass function (pdf/pmf) $\prob(\relvec)$. To simplify the discussion, we assume it to be discrete and memoryless.\footnote{For the remainder of the article, note that the domain of all \rv s $\dom$ may be either discrete or continuous. Further, we note that the definition of entropy for discrete and continuous \rv s differs. For example, the differential entropy of continuous \rv s may be negative whereas the entropy of discrete \rv s is always positive~\cite{simeone2018brief}. Without loss of generality, we will thus assume all \rv s either to be discrete or to be continuous. In this work, we avoid notational clutter by using the expected value operator: Replacing the integral by summation over discrete \rv s, the equations are also valid for discrete \rv s and vice versa.}

Then, a semantic channel modeled by conditional distribution $\prob(\semvec|\relvec)$ generates an observation or source signal, a \rv\ $\semvec\in\semset^{\Nsemvec\times 1}$, that enters the communication system. Compared to~\cite{bao_towards_2011} where the semantic channel is the transmission system, we consider probabilistic semantic channels $\prob(\semvec|\relvec)$ using the definition from~\cite{beck_semantic_2023}. We refer the reader to~\cite{beck_semantic_2023} for an example of what these \rv s may look like.

\subsubsection{Semantic Channel Encoding}

Our challenge is to encode the source $\semvec$ onto the transmit signal $\xvec\in\xset^{\ntx\times 1}$ (see Fig.~\ref{fig1:com_system_small}) for efficient and reliable semantic transmission through the physical communication channel $\prob(\yvec|\xvec)$, where $\yvec\in\yset^{\Nrx\times 1}$ is the received signal vector, such that the semantic \rv\ $\relvec$ at a recipient is best preserved~\cite{beck_semantic_2023}. We parametrize the encoder $\prob_{\txpars}(\xvec|\semvec)$ by a parameter vector $\txpars\in\realnum^{\Ntxpars\times 1}$ and assume $\prob_{\txpars}(\xvec|\semvec)$ to be deterministic in communications with $\prob_{\txpars}(\xvec|\semvec)=\dirac(\xvec-\encdet)$ and encoder function $\encdet$. In summary, we bring the semantic source $\relvec$ to the context of communications by considering the complete Markov chain $\relvec \leftrightarrow \semvec \leftrightarrow \xvec \leftrightarrow \yvec$ in contrast to~\cite{bao_towards_2011}.

In classic Shannon design, the posterior $\prob_{\txpars}(\semvec|\yvec)$ is processed to recover the observation $\semvec$ as accurately as possible at the receiver side. Instead, we recover semantics $\relvec$ processing $\prob_{\txpars}(\relvec|\yvec)$: %
Since the entropy $\HE[\relvec]=\evalnb[\relvec\sim\prob(\relvec)]{-\ln \prob(\relvec)}$ of the semantic \rv\ $\relvec$ is expected to be less or equal to the entropy $\HE[\semvec]$ of the source $\semvec$, i.e., $\HE[\relvec]\le\HE[\semvec]$, we can compress by transmitting the semantic \rv\ $\relvec$. There, $\eval[\xvec\sim \prob(\xvec)]{\funcrtrick(\xvec)}$ denotes the expected value of $\funcrtrick(\xvec)$ w.r.t. both discrete or continuous \rv s $\xvec$.

\subsection{Semantic Communication Design}

Now, we revisit our two design approaches from~\cite{beck_semantic_2023}.

\subsubsection{InfoMax Principle}
\label{sec:infomaxlearning}

First, we like to find the encoder $\prob_{\txpars}(\xvec|\semvec)$ that maps $\semvec$ to a representation $\yvec$ such that most information of the relevant \rv\ $\relvec$ is included in $\yvec$, i.e., we maximize the Mutual Information (MI) $\mi[\txpars]{\relvec;\yvec}$ w.r.t. $\prob_{\txpars}(\xvec|\semvec)$:
\begin{align}
	  & \argmax[\prob_{\txpars}(\xvec|\semvec)]\mi[\txpars]{\relvec;\yvec} \label{eq:infomax}                                                                                          \\
	= & \argmax[\prob_{\txpars}(\xvec|\semvec)]\eval[\relvec,\yvec\sim\prob_{\txpars}(\relvec,\yvec)]{\ln \frac{\prob_{\txpars}(\relvec,\yvec)}{\prob(\relvec)\prob_{\txpars}(\yvec)}} \\
	= & \argmax[\prob_{\txpars}(\xvec|\semvec)]\HE[\relvec] - \HE[\prob_{\txpars}(\relvec,\yvec), \prob_{\txpars}(\relvec|\yvec)] \label{eq:infomax_connection}                        \\
	= & \argmax[\prob_{\txpars}(\xvec|\semvec)]\eval[\relvec,\yvec\sim\prob_{\txpars}(\relvec,\yvec)]{\ln \prob_{\txpars}(\relvec|\yvec)} \eqpoint \label{eq:infomax2}
\end{align}
There, $\HE[\prob(\xvec), \aprob(\xvec)]=\evalnb[\xvec\sim\prob(\xvec)]{-\ln \aprob(\xvec)}$ is the cross entropy between two pdfs/pmfs $\prob(\xvec)$ and $\aprob(\xvec)$.

If the posterior $\prob_{\txpars}(\relvec|\yvec)$ in~\eqref{eq:infomax2} is intractable to compute, we can replace it with a variational distribution $\aprob_{\rxpars}(\relvec|\yvec)$ with parameters $\rxpars\in\realnum^{\Nrxpars\times 1}$, i.e., the semantic decoder in Fig.~\ref{fig1:com_system_small}. Then, we can define a MI Lower BOund (MILBO)~\cite{beck_semantic_2023}:
\begin{align}
	\mi[\txpars]{\relvec;\yvec} & \geq \eval[\relvec,\yvec\sim\prob_{\txpars}(\relvec,\yvec)]{\ln \aprob_{\rxpars}(\relvec|\yvec)} \label{eq:milbo}        \\
	                            & = -\eval[\yvec\sim\prob(\yvec)]{\HE[\prob_{\txpars}(\relvec|\yvec), \aprob_{\rxpars}(\relvec|\yvec)]} \label{eq:ce_loss} \\
	                            & = -\ceparams  \eqpoint \label{eq:ce_loss_opt}
\end{align}
Now, we can learn optimal parametrizations $\txpars$ and $\rxpars$ of the transmitter discriminative model $\prob_{\txpars}(\xvec|\semvec)$ and of the variational receiver posterior $\aprob_{\rxpars}(\relvec|\yvec)$ by minimizing the amortized cross entropy $\ceparams$ in~\eqref{eq:ce_loss}, i.e., marginalized across received signals $\yvec$~\cite{beck_semantic_2023}. The encoder can be seen by rewriting:
\begin{align}
	\ceparams & = \eval[\semvec,\xvec,\yvec,\relvec\sim \prob_{\txpars}(\semvec,\xvec,\yvec,\relvec)]{-\ln \aprob_{\rxpars}(\relvec|\yvec)} \label{eq:relvar_infomax}                                           \\
	          & = \eval[\semvec,\relvec\sim\prob(\semvec,\relvec)]{\eval[\xvec\sim\prob_{\txpars}(\xvec|\semvec)]{\eval[\yvec\sim\prob(\yvec|\xvec)]{-\ln \aprob_{\rxpars}(\relvec|\yvec)}}} \eqpoint \nonumber
\end{align}
The idea is to solve~\eqref{eq:relvar_infomax} by AEs or -- in this article -- RL. Thus, we use DNNs for the design of both encoder $\prob_{\txpars}(\xvec|\semvec)$ and decoder $\aprob_{\rxpars}(\relvec|\yvec)$~\cite{xie_deep_2021}.

Note that in our semantic problem~\eqref{eq:infomax} or~\eqref{eq:relvar_infomax}, we do not auto encode the hidden $\relvec$ or $\semvec$ as in~\cite{xie_deep_2021} itself, but encode $\semvec$ to obtain $\relvec$ by decoding. This means our interpretation of semantic information and its recovery deviates from literature: We define semantics $\relvec$ explicitly compared to, e.g.,~\cite{xie_deep_2021}, that optimizes on $\semvec$ and then measures semantic similarity w.r.t. its estimate $\semhvec$ explicitly by some semantic metric $\Lossf(\semvec,\semhvec)$.

\subsubsection{Information Bottleneck View}

Further, introducing a constraint on the information rate in~\eqref{eq:infomax}, we can formulate an Information Bottleneck (IB) optimization problem~\cite{calvanese_strinati_6g_2021}, where we like to maximize the relevant information $\mi[\txpars]{\relvec;\yvec}$ subject to the constraint to limit the compression rate $\mi[\txpars]{\semvec;\yvec}$ to a maximum information rate $\mic$:
\begin{align}
	 & \argmax[\prob_{\txpars}(\xvec|\semvec)]\mi[\txpars]{\relvec;\yvec} \st \mi[\txpars]{\semvec;\yvec}\le \mic  \eqpoint \label{eq:ibm}
\end{align}
In this article, we set constraint $\mic$ by fixing $\ntx$ since then an upper bound on $\mi[\txpars]{\semvec;\yvec}$ grows as shown in~\cite{beck_semantic_2023}. With fixed constraint $\mic$, we then need to maximize the relevant information $\mi[\txpars]{\relvec;\yvec}$. As in the InfoMax problem, we can exploit the MILBO to use the amortized cross entropy $\ceparams$ in~\eqref{eq:relvar_infomax} as the optimization criterion.

\section{Stochastic Policy Gradient-based Reinforcement Learning}
\label{sec:3}

If calculating the expected value of the amortized cross entropy $\ceparams$ in~\eqref{eq:relvar_infomax} is analytically or computationally intractable as typical with DNNs, we can approximate it using Monte Carlo sampling techniques with $\Nsamp$ samples $\lcb\relvecreal_{\indi},\semvecreal_{\indi},\xvecreal_{\indi},\yvecreal_{\indi}\rcb_{i=1}^{\Nsamp}$.

\subsection{Stochastic Gradient Descent-based Optimization}

For Stochastic Gradient Descent (SGD) - based optimization, the gradient w.r.t. $\rxpars$ can then be calculated by
\begin{align}
	\frac{\partial\ceparams}{\partial\rxpars}%
	=       & -\eval[\relvec,\semvec,\yvec\sim\prob_{\txpars}(\yvec|\semvec)\prob(\semvec|\relvec)\prob(\relvec)]{\frac{\partial \ln\aprob_{\rxpars}(\relvec|\yvec)}{\partial\rxpars}} \\
	\approx & \frac{1}{\Nsamp}\summ[\Nsamp]{\indi=1}\frac{\partial\lbb-\ln \aprob_{\rxpars}(\relvecreal_{\indi}|\yvecreal_{\indi})\rbb}{\partial \rxpars} \label{eq:sgdrxpars}
\end{align}
with $\Nsamp$ being equal to the batch size $\Nb$ and by application of the backpropagation algorithm in Automatic Differentiation Frameworks (ADF), e.g., TensorFlow or PyTorch.

\subsubsection{Reinforce Gradient}
Computing the gradient w.r.t. $\txpars$ is not straightforward since we sample w.r.t. the distribution $\prob_{\txpars}(\yvec|\semvec)$ dependent on $\txpars$~\cite{simeone2018brief}. For continuous-valued $\yvec$ and using the log-trick $\frac{\partial \ln \prob_{\txpars}(\yvec|\semvec)}{\partial\txpars}=\frac{\partial\prob_{\txpars}(\yvec|\semvec)}{\partial\txpars}/\prob_{\txpars}(\yvec|\semvec)$, we derive:
\begin{align}
	        & \frac{\partial\ceparams}{\partial\txpars} \nonumber                                                                                                                                                                                                                                                                                                       \\
	=       & -\frac{\partial}{\partial\txpars}\eval[\relvec,\semvec,\yvec\sim\prob_{\txpars}(\yvec|\semvec)\prob(\semvec,\relvec)]{\ln \aprob_{\rxpars}(\relvec|\yvec)} \label{eq:ce_grad}                                                                                                                                                                             \\
	=       & -\E_{\relvec,\semvec\sim\prob(\semvec,\relvec)}\negthinspace\bigg[\int_{\yset^{\Nrx}} \underbrace{\frac{\partial\prob_{\txpars}(\yvec|\semvec)}{\partial\txpars}}_{=\prob_{\txpars}(\yvec|\semvec)\cdot\frac{\partial\ln \prob_{\txpars}(\yvec|\semvec)}{\partial\txpars}} \cdot \ln \aprob_{\rxpars}(\relvec|\yvec)\; \D\yvec\bigg] \label{eq:logtrick1} \\
	=       & -\eval[\relvec,\semvec,\yvec\sim\prob_{\txpars}(\yvec|\semvec)\prob(\semvec,\relvec)]{\frac{\partial\ln \prob_{\txpars}(\yvec|\semvec)}{\partial\txpars}\cdot \ln \aprob_{\rxpars}(\relvec|\yvec)} \label{eq:reinforce_grad2}                                                                                                                             \\
	\approx & -\frac{1}{\Nsamp}\summ[\Nsamp]{\indi=1}\frac{\partial\ln \prob_{\txpars}(\yvecreal_{\indi}|\semvecreal_{\indi})}{\partial\txpars}\cdot \ln \aprob_{\rxpars}(\relvecreal_{\indi}|\yvecreal_{\indi}) \label{eq:reinforce} \eqpoint
\end{align}
We arrive at the same result with discrete \rv s $\yvec$ replacing the integral in~\eqref{eq:logtrick1} by a sum. The Monte Carlo approximation~\eqref{eq:reinforce} is the \reinforce\ gradient w.r.t. $\txpars$~\cite{simeone2018brief}. This estimate has high variance since we sample w.r.t. the distribution $\prob_{\txpars}(\yvec|\semvec)$ dependent on $\txpars$.

\subsubsection{Reparametrization Trick}

Leveraging the direct relationship between $\txpars$ and $\yvec$ in $\ln \aprob_{\rxpars}(\relvec|\yvec)$ can help reduce the estimator's high variance. Typically, e.g., in Variational AEs (VAE), the reparametrization trick is used to achieve this~\cite{simeone2018brief}. Here we can apply it if we can decompose the latent variable $\yvec\sim\prob_{\txpars}(\yvec|\semvec)$ into a differentiable function $\yvec=\funcrtrick_{\txpars}(\semvec,\noisevec)$ and a \rv\ $\noisevec\sim \prob(\noisevec)$ independent of $\txpars$. Fortunately, the typical forward model of a communication system $\prob_{\txpars}(\yvec|\semvec)$ fulfills this criterion. Assuming a deterministic (DNN) encoder $\xvec=\encdet$ and additive noise $\noisevec$ with covariance $\covariance$, we can thus rewrite $\yvec$ into $\funcrtrick_{\txpars}(\semvec, \noisevec)=\encdet+\covariance^{1/2}\cdot \noisevec$ and accordingly the amortized cross entropy gradient~\eqref{eq:ce_grad} into:
\begin{align}
	\frac{\partial\ceparams}{\partial\txpars} %
	=       & -\eval[\relvec,\semvec,\noisevec\sim\prob(\noisevec)\prob(\semvec|\relvec)\prob(\relvec)]{\frac{\partial \funcrtrick_{\txpars}(\semvec,\noisevec)}{\partial\txpars} \cdot \frac{\partial \ln \aprob_{\rxpars}(\relvec|\yvec)}{\partial\yvec}}                                                                                            \\
	\approx & -\frac{1}{\Nsamp}\summ[\Nsamp]{\indi=1}\frac{\partial \funcrtrick_{\txpars}(\semvecreal_{\indi},\noisevecreal_{\indi})}{\partial\txpars}\cdot \frac{\partial \ln \aprob_{\rxpars}(\relvecreal_{\indi}|\yvec)}{\partial\yvec}\bigg|_{\yvec=\funcrtrick_{\txpars}(\semvecreal_{\indi}, \noisevecreal_{\indi})} \label{eq:reparam} \eqpoint
\end{align}
The trick can be easily implemented in ADFs by adding a noise layer after function $\xvec=\encdet$, typically used for regularization in ML literature. Then, our loss function~\eqref{eq:relvar_infomax} is the empirical cross entropy:
\begin{align}
	\ceparams%
	\approx -\frac{1}{\Nsamp}\summ[\Nsamp]{\indi=1}\ln \aprob_{\rxpars}(\relvecreal_{\indi}|\yvecreal_{\indi}=\funcrtrick_{\txpars}(\semvecreal_{\indi}, \noisevecreal_{\indi})) \eqpoint \label{eq:mcloss}
\end{align}
This allows for joint learning of both $\txpars$ and $\rxpars$, as demonstrated in recent works~\cite{oshea_introduction_2017,beck_semantic_2023}, treating unsupervised optimization of AEs and \sinfoni\ as a supervised learning problem.

\subsection{Stochastic Policy Gradient}

We note that optimization of encoder and decoder with both gradients~\eqref{eq:reinforce} or~\eqref{eq:reparam} requires model-awareness, i.e., a known and differentiable forward model $\prob_{\txpars}(\yvec|\semvec)$. But the gradient
\begin{align}
	\frac{\partial \ln\prob_{\txpars}(\yvec|\semvec)}{\partial\txpars}=\frac{\partial \encdet}{\partial\txpars} \cdot \frac{\partial \prob(\yvec|\xvec)}{\partial\xvec}\cdot \frac{\partial \ln\prob(\yvec|\xvec)}{\partial\prob(\yvec|\xvec)} \label{eq:non_differentiable_channel} %
\end{align}
with deterministic encoder $\xvec=\encdet$ may not be computable, as the channel model $\prob(\yvec|\xvec)$ could be non-differentiable or unknown without any channel estimate. Further, in practice, the transmitter and receiver are separated at different locations and have at most a rudimentary feedback link, requiring independent optimization w.r.t. $\txpars$ and $\rxpars$: The transmitter does not know $\aprob_{\rxpars}(\relvec|\yvec)$ and the receiver $\prob_{\txpars}(\xvec|\semvec)$, vice versa.

To tackle these challenges in gradient computation, we now introduce a stochastic policy $\prob_{\txpars}(\xvec|\semvec)\neq\dirac(\xvec-\encdet)$ that fulfills the reparametrization property:
\begin{align}
	\frac{\partial \ceparams}{\partial\txpars} %
	=       & -\frac{\partial}{\partial\txpars}\eval[\relvec,\semvec,\xvec,\yvec\sim\prob(\yvec|\xvec)\prob_{\txpars}(\xvec|\semvec)\prob(\semvec,\relvec)]{\ln \aprob_{\rxpars}(\relvec|\yvec)}                                                                                        \\
	=       & -\E_{\relvec,\semvec\sim\prob(\semvec,\relvec)}\bigg[ \int_{\xset^{\ntx}}  \underbrace{\frac{\partial\prob_{\txpars}(\xvec|\semvec)}{\partial\txpars}}_{=\prob_{\txpars}(\xvec|\semvec)\cdot\frac{\partial\ln \prob_{\txpars}(\xvec|\semvec)}{\partial\txpars}} \nonumber \\
	        & \cdot \eval[\yvec\sim\prob(\yvec|\xvec)]{\ln \aprob_{\rxpars}(\relvec|\yvec)}\; \D\xvec \bigg] \label{eq:logtrick2}                                                                                                                                                       \\
	=       & -\eval[\relvec,\semvec,\xvec,\yvec\sim\prob_{\txpars}(\relvec,\semvec,\xvec,\yvec)]{\frac{\partial\ln \prob_{\txpars}(\xvec|\semvec)}{\partial\txpars}\cdot \ln \aprob_{\rxpars}(\relvec|\yvec)} \label{eq:policy_grad}                                                   \\
	\approx & \frac{1}{\Nsamp}\summ[\Nsamp]{\indi=1}\frac{\partial\ln \prob_{\txpars}(\xvecreal_{\indi}|\semvecreal_{\indi})}{\partial\txpars}\cdot \lbb-\ln \aprob_{\rxpars}(\relvecreal_{\indi}|\yvecreal_{\indi})\rbb \label{eq:reinforce2} \eqpoint
\end{align}
Again the log-trick is applied in~\eqref{eq:logtrick2} to arrive in~\eqref{eq:policy_grad} and the results hold for discrete \rv s $\xvec$. Most importantly,~\eqref{eq:policy_grad} is the policy gradient and the derivation is equivalent to the Stochastic Policy Gradient (SPG) theorem, a fundamental result of continuous-action RL~\cite{silver_deterministic_2014}. For integration into ADFs, usually, an objective function whose gradient is the Monte Carlo policy gradient estimator of~\eqref{eq:policy_grad}, i.e., the \reinforce\ gradient~\eqref{eq:reinforce2}, is constructed:%
\begin{align}
	\spgobj=\frac{1}{\Nsamp}\summ[\Nsamp]{\indi=1}\ln \prob_{\txpars}(\xvecreal_{\indi}|\semvecreal_{\indi})\cdot \lbb-\ln \aprob_{\rxpars}(\relvecreal_{\indi}|\yvecreal_{\indi})\rbb \eqpoint \label{eq:policy_grad_obj}
\end{align}
With objective~\eqref{eq:policy_grad_obj} or \reinforce\ gradient~\eqref{eq:reinforce2}, we can finally optimize $\ceparams$ w.r.t. $\txpars$, since we can sample $\lcb\relvec,\semvec,\xvec,\yvec\rcb\sim\prob_{\txpars}(\relvec,\semvec,\xvec,\yvec)$ and compute $\frac{\partial\ln \prob_{\txpars}(\xvecreal_{\indi}|\semvecreal_{\indi})}{\partial\txpars}$ at the transmitter and $-\ln \aprob_{\rxpars}(\relvecreal_{\indi}|\yvecreal_{\indi})$ being equal to the per-sample cross entropy at the receiver.

Note that $\semtrain$ and $\xtrain$ only have to be known at the transmitter and both $\reltrain$ and $\ytrain$ at the receiver, respectively. This enables the separation or spatial distribution of transmitter and receiver when the following conditions are met:
\begin{itemize}
	\item Only an a priori known pilot sequence $\pilotset=\lcb\relvecreal_{\indi},\semvecreal_{\indi}\rcb_{i=1}^{\Npilot}$ of size $\Npilot$ is required. This sequence translates into the training set $\trainingset=\lcb\relvecreal_{\indi},\semvecreal_{\indi},\xvecreal_{\indi},\yvecreal_{\indi}\rcb_{i=1}^{\Ntrain}$ of size $\Ntrain$ which is divided into batches of size $\Nb$ for SGD-based optimization.
	\item Moreover, we require a feedback link to transmit the per-sample cross entropy $-\ln \aprob_{\rxpars}(\reltrain|\ytrain)$ to the encoder. This term can be interpreted as a reward or critic known from RL~\cite{silver_deterministic_2014}. Accordingly, the transmitter can be seen as an actor with a policy $\prob_{\txpars}(\xvec|\semvec)$. The best continuous action/policy is then learned by optimization w.r.t. these rewards.
\end{itemize}

\subsubsection{Stochastic Policy}

Introducing a stochastic policy means we need to add a probabilistic sampler/explorer function $\prob(\xvec|\xvecexpl)$ to the encoder as shown in Fig.~\ref{fig:com_system_RL}. Replacing $\prob(\yvec|\semvec)$ and $\prob(\yvec|\xvec)$ by $\prob(\xvec|\semvec)$ and $\prob(\xvec|\xvecexpl)$ in~\eqref{eq:non_differentiable_channel} and applying the result to \eqref{eq:reinforce2}, we derive that this function needs to be differentiable. If the encoder output, i.e., the action space, is continuous with $\xset=\realnum$, we can achieve this using for example a Gaussian policy, i.e., a multivariate Gaussian pdf
\begin{equation}
	\prob(\xvec|\xvecexpl)=\prob\lpa\xvec|\xvecexpl,\perstd^2\rpa=\normdis\lpa (1-\perstd^2)^{1/2}\cdot\xvecexpl,\perstd^2\cdot\eye\rpa \label{eq:gaussian_policy}
\end{equation}
with exploration variance $\perstd^2\in(0,1)$ where scaling of the mean $\xvecexpl=\encdet$ is done to ensure the conservation of average energy. Furthermore, the Gaussian policy offers the benefit of simplicity in parametrization, requiring tuning of only two pdf parameters. Hence, we employ it in our numerical experiments. For discrete action spaces $\xset^{\ntx\times 1}$, a continuous differentiable relaxation such as the Gumbel Softmax is required \cite{beck_cmdnet_2021}.

In the special case~$\perstd^2\rightarrow 0$, the Gaussian policy $\prob(\xvec|\xvecexpl,\perstd^2)$ approaches a deterministic policy. In~\cite{aoudia_model-free_2019}, the authors show that the true channel gradient $\frac{\partial}{\partial\xvec} \prob(\yvec|\xvec)$ is then perfectly approximated. However, using a near-deterministic policy leads in their experiments to high variance of the gradient estimate~\eqref{eq:reinforce2} resulting in slow convergence. To compensate for this effect, we require a much larger and computationally expensive batch size $\Nsamp=\Nb$. From the view of RL, using a stochastic policy with $\perstd^2\neq 0$ enables the exploration of the set of possible actions.

\subsection{Alternating RL-based Training}

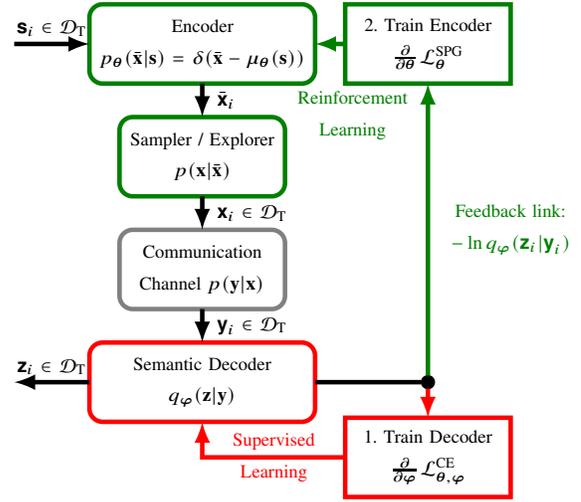
\begin{figure}[!t]
	\centerline{\input{./TikZ/com_system_model_basic_RL.tikz}}
	\caption{Optimization procedure of a semantic encoder and decoder without a differentiable channel model: 1. Train the decoder supervised based on the training sequence and updated encoder but without sampler. 2. Encoder explores transmit signals $\xtrain$ and improves its policy according to the decoder reward feedback. 3. Alternate between both steps until convergence.}
	\label{fig:com_system_RL}
\end{figure}

After introducing the SPG, we now derive an optimization procedure akin to~\cite{aoudia_model-free_2019} for the whole semantic communication system. It does not require any channel model but a fixed pilot, i.e., training, sequence and a feedback link. Further, it enables separation of encoder and decoder. We show it in Fig.~\ref{fig:com_system_RL}:
\begin{enumerate}
	\item We note that according to~\eqref{eq:sgdrxpars} decoder optimization reduces to supervised learning w.r.t. $\ceparams$ and $\rxpars$ at the receiver side. Thus, in the first step, we train the decoder based on the training sequence and updated encoder, but without sampler/explorer ($\perstd^2=0$).
	\item Second, the encoder explores with transmit signals $\xtrain$. It is optimized based on the policy gradient of $\spgobj$ and the reward $-\ln \aprob_{\rxpars}(\reltrain|\ytrain)$ that the decoder feeds back.
	\item We alternate between the first and second training steps until convergence. Note that we can use one or multiple SGD steps and batches for each alternating training step, respectively.
\end{enumerate}
Reminiscent of the RL fashion of the stochastic policy optimization of Semantic INFOrmation traNsmission and recoverY~\cite{beck_semantic_2023}, we name this approach \sinfonirl. Finally, we have derived the SPG for semantic communication starting from the InfoMax problem \eqref{eq:infomax}. Replacing $\mi[\txpars]{\relvec;\yvec}$ by $\mi[\txpars]{\semvec;\yvec}$, this result also holds more general for classic communications.

\section{Example of Model-free Semantic Recovery}
\label{sec:4}

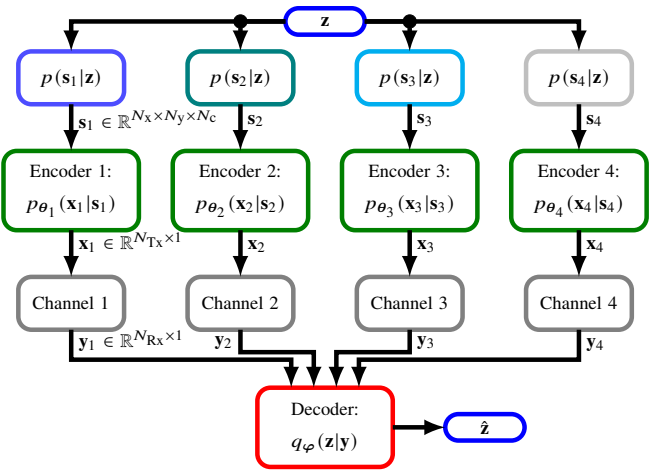
\begin{figure}[!t]
	\centerline{\input{./TikZ/distributed_resnet_v5_small_RL.tikz}}
	\caption{\sinfonirl\ scenario: Four distributed agents extract features for rate-efficient transmission to a decoder that extracts semantics.}
	\label{fig:resnet_distr}
\end{figure}

To evaluate the proposed model-free optimization approach \sinfonirl, we use the numerical example of distributed image classification with \sinfoni\ from~\cite{beck_semantic_2023} shown in Fig.~\ref{fig:resnet_distr}. Thus, we will now assume the hidden semantic \rv\ to be a one-hot vector $\relvec\in \{0,1\}^{\Nclass\times 1}$ representing one of $\Nclass$ image classes. Then, each of the four agents observes its image, i.e., the observation $\semvec_{\indi}\sim\prob(\semvec_{\indi}|\relvec)$ with $\indi=1,\dots,4$, through a semantic channel, being generated by the same semantic \rv\ $\relvec$ and thus belonging to the same class. Based on these images, a central unit shall extract semantics, i.e., perform classification.

We propose to optimize the four encoders $\prob_{\txpars_{\indi}}(\xvec_{\indi}|\semvec_{\indi})$ \textbf{jointly} with a decoder $\aprob_{\rxpars}(\relvec|\yvec=[\yvec_1,\yvec_2,\yvec_3,\yvec_4]^{\trapo})$ w.r.t. cross entropy~\eqref{eq:relvar_infomax} of the semantic labels (see Fig.~\ref{fig:resnet_distr}). Hence, we maximize the system's overall semantic measure, i.e., classification accuracy.

To show the basic working principle and ease implementation, we use the grayscale MNIST and colored CIFAR10 datasets with $\Nclass=10$ image classes~\cite{beck_semantic_2023}. We assume that the semantic channel generates an image that we divide into four equally sized quadrants and each agent observes one quadrant $\semvec_{\indi}\in \realnum^{\Nimx\times\Nimy\times\Nimc}$ where $\Nimx$ and $\Nimy$ is the number of image pixels in the x- and y-dimension, respectively, and $\Nimc$ is the color channel number.

\subsection{Distributed \sinfoni\ Approach}

For the design of \sinfoni, we rely on the powerful DNN approach ResNet for feature extraction~\cite{beck_semantic_2023}. We use the pre-activation version of ResNet without bottlenecks implemented for CIFAR10 classification. In Tab.~\ref{tab2:com_sem}, we show its structure modified for the distributed scenario from Fig.~\ref{fig:resnet_distr}. There, ResNetBlock is the basic building block of the ResNet architecture. Each block consists of multiple residual units (res. un.) and we use $2$ for the MNIST and $3$ for the CIFAR10 dataset. For further implementation details, we refer the reader to the original work~\cite{beck_semantic_2023} and our source code~\cite{Beck_Semantic_INFOrmation_traNsmission_2023}.

\begin{table}[!t]
	\renewcommand{\arraystretch}{1.3}
	\caption{\sinfonirl\ - DNN architecture for image example.}
	\label{tab2:com_sem}
	\centering
	\begin{tabular}{lll}
		\hline
		Component         & Layer                      & Dimension                      \\
		\hline
		Input             & Image (MNIST, CIFAR10)     & $(14, 14, 1)$, $(16, 16, 3)$   \\
		\hline
		$4\times$         & Conv2D                     & $(14, 14, 14)$, $(16, 16, 16)$ \\
		Feature           & ResNetBlock (2/3 res. un.) & $(14, 14, 14)$, $(16, 16, 16)$ \\
		Extractor         & ResNetBlock (2/3 res. un.) & $(7, 7, 28)$, $(8, 8, 32)$     \\
		                  & ResNetBlock (2/3 res. un.) & $(4, 4, 56)$, $(4, 4, 64)$     \\
		                  & Batch Normalization        & $(4, 4, 56)$, $(4, 4, 64)$     \\
		                  & ReLU activation            & $(4, 4, 56)$, $(4, 4, 64)$     \\
		                  & GlobalAvgPool2D            & $(56)$, $(64)$                 \\
		\hline
		$4\times$ Tx      & ReLU                       & $\ntx$                         \\
		                  & Linear                     & $\ntx$                         \\
		                  & Normalization (dim.)       & $\ntx$                         \\
		\hline
		$4\times$ Sampler & AWGN + Normalization       & $\ntx$                         \\
		\hline
		$4\times$ Channel & AWGN                       & $\ntx$                         \\
		\hline
		Rx                & ReLU ($4\times$ shared)    & $(2, 2, \nrx)$                 \\
		                  & GlobalAvgPool2D            & $\nrx$                         \\
		\hline
		Classifier        & Softmax                    & $\Nclass=10$                   \\
		\hline
	\end{tabular}
\end{table}

Our key idea here is to modify ResNet w.r.t. the communication task by splitting it where a low-bandwidth representation of semantic information is present. Therefore, we aim to transmit each agent's local features of length $\Nfeat$ provided by the Feature Extractors in Tab.~\ref{tab2:com_sem} instead of all sub-images $\semvec_{\indi}$ and add the component Tx to encode the features into $\xvec_{\indi}\in\realnum^{\ntx\times 1}$ for transmission through the wireless channel (see Fig.~\ref{fig:resnet_distr}).
We note that $\xvec_{\indi}\in\realnum^{\ntx\times 1}$ is analog and that the output dimension $\ntx$ defines the number of channel uses per agent and thus information rate. To limit the transmit power to one, we constrain the Tx Linear layer output by the norm along the training batch or the encode vector dimension (dim.). %

For \sinfonirl, we add a Gaussian Sampler~\eqref{eq:gaussian_policy} after the Tx output compared to~\cite{beck_semantic_2023}. Further, we assume all agents and the Rx module to share a training set $\trainingset$ and a perfect reward feedback link from the Rx module to all agents.

At the receiver side, we use a single Rx module only with shared DNN layers of width $\nrx$ and parameters $\rxpars_{\txt{Rx}}$ for all inputs $\yvec_{\indi}$~\cite{beck_semantic_2023}.
Based on an aggregation of the four Rx outputs, a softmax layer with $\Nclass=10$ units finally computes class probabilities $\aprob_{\rxpars}(\relvec|\yvec)$ whose maximum is the maximum a posteriori estimate $\relvecest$.

\subsection{Optimization Details}
\label{sec:43}

We evaluate \sinfonirl\ in TensorFlow 2 on the MNIST and CIFAR10 datasets with training set $\trainingset$~\cite{Beck_Semantic_INFOrmation_traNsmission_2023}. For cross-entropy loss minimization, we use the gradient approximations from Sec.~\ref{sec:3} and the SGD-variant Adam with a batch size of $\Nb=500$. We add $\lnorm[2]$-regularization with a weight decay of $0.0001$. To optimize the transceiver for a wider SNR range, we choose the SNR to be uniformly distributed within $[-4,6]$ dB where $\snr=1/\noisestd^2$ with noise variance $\noisestd^2$. We set $\nrx=\Nfeat$ as default and refer to~\cite{beck_semantic_2023, Beck_Semantic_INFOrmation_traNsmission_2023} for more implementation details. In the following, we compare the performance of\footnote{It is not straightforward to compare the approach from \cite{lu_reinforcement_2022} with \sinfonirl\ as different models were investigated. We leave a detailed comparison with other approaches from the literature for future work.}: %
\begin{itemize}
	\item \textbf{\centralimagecomm:} Digital transmission baseline from~\cite{beck_semantic_2023} with capacity achieving LDPC code and ResNet classifier.
	\item \textbf{\sinfoni:} The distributed \sinfoni\ design from~\cite{beck_semantic_2023} trained \textit{model-aware} as one DNN with channel noise layer using the reparametrization trick~\eqref{eq:reparam} to approximate the gradients. We train for $\Ne=100$ epochs with the MNIST dataset.
	\item \textbf{\sinfonirl:} New approach trained \textit{model-free} via RL as shown in Fig.~\ref{fig:com_system_RL} using SPG~\eqref{eq:reinforce2}. We alternate between $10$ decoder and encoder optimization steps. Note that one decoder and encoder step amounts to one iteration of the model-aware approach where the encoder and decoder are optimized jointly. Hence, for a fair comparison, we divide the number of alternating iterations or epochs $\Ne$ of the SPG approach by $2$. We choose $\Ne=3000$ and add $\Nerx=600$ epochs of receiver fine-tuning at the end~\cite{aoudia_model-free_2019}. To decrease the SPG estimator variance, we choose a rather high exploration variance $\perstd^2=0.15$.
	\item[\textbf{-}] \textbf{\perfcomm:} \sinfoni\ trained with perfect communication links without Tx and Rx modules, but with Tx normalization. Thus, the plain power-constrained features are transmitted with $\ntx=56$ or $64$ channel uses. It serves as the benchmark, as it indicates the maximum performance of the distributed design.
	\item[\textbf{-}] \textbf{\txrx\ $\bog{\ntx}$:} Default \sinfoni\ from Tab.~\ref{tab2:com_sem} trained with Tx and Rx module and $\ntx$ channel uses.
\end{itemize}

\subsection{Numerical Results}

To measure semantic transmission quality, we use classification error rate on semantic \rv\ $\relvec$ and normalize the SNR by the spectral efficiency $\speceff=\Nfeat/\ntx$~\cite{beck_semantic_2023}.

\subsubsection{MNIST dataset}

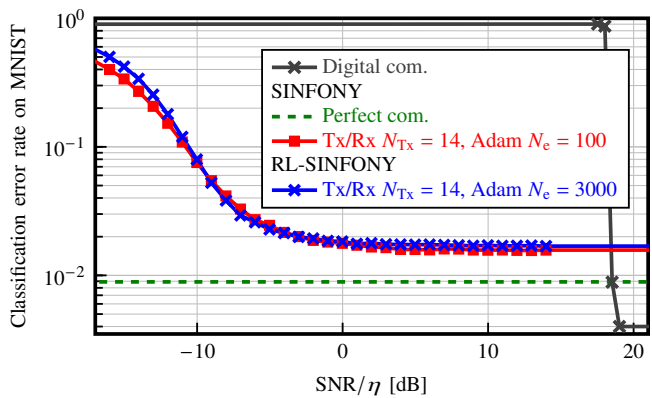
\begin{figure}[!t]
	\centerline{\input{./TikZ/comsem_image_mnist_snr-4_6_RL_normSNR.tikz}}
	\caption{Comparison of the classification error rate of \sinfonirl\ and \sinfoni\ with $\ntx=14$ on MNIST as a function of normalized SNR.}
	\label{fig:sinfony_mnist_RL}
\end{figure}

The numerical results of our proposed approach \sinfonirl\ on the MNIST validation dataset are shown in Fig.~\ref{fig:sinfony_mnist_RL}. We observe that both approaches \sinfonirl\ and \sinfoni\ with Tx/Rx module approach the benchmark with ideal links (\sinfoniperfcomm) at high SNR and beat \centralimagecomm\ w.r.t. communication efficiency. Notably, both curves are very close to each other, i.e., the performance gap after training is minor. This means training of \sinfonirl\ converged successfully. Note that \centralimagecomm\ classifies the entire image at once and thus outperforms at high SNR~\cite{beck_semantic_2023}.

\subsubsection{Convergence Rate}

\begin{figure}[!t]
	\centerline{\input{./TikZ/comsem_image_mnist_snr-4_6_RL_convergence.tikz}}
	\caption{Comparison of training convergence between \sinfonirl\ and \sinfoni\ with $\ntx=14$ in terms of the cross-entropy loss on MNIST averaged over $10$ runs as a function of training epochs $\Ne$.}
	\label{fig:sinfony_convergence_RL}
	\vspace{-0.5cm}
\end{figure}
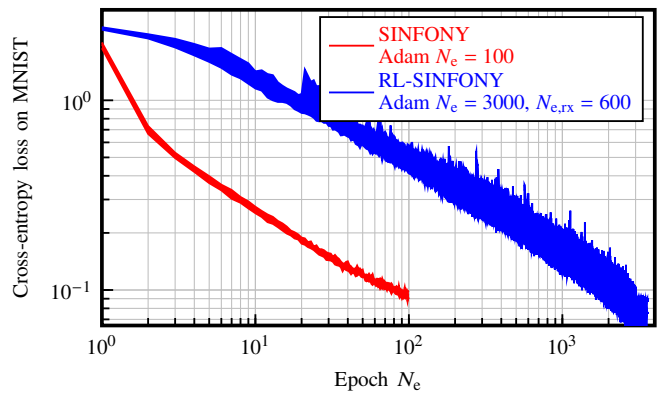

Since the number of training epochs required to achieve the same performance deviates significantly with $\Ne+\Nerx=3000+600=3600$ compared to $\Ne=100$, we take a closer look at training convergence in terms of the cross-entropy loss shown in Fig.~\ref{fig:sinfony_convergence_RL}. We averaged the loss over $10$ training runs and illustrate the interval between the maximum and minimum loss value using shaded areas. To reach the same loss, we require more than $10$ times more epochs with \sinfonirl\ compared to \sinfoni. The reason for the decreased convergence is the increased variance of the \reinforce\ gradient~\eqref{eq:reinforce2} compared to the reparametrization trick gradient~\eqref{eq:reparam}. Also, we attribute the increased variance in training losses (blue-shaded area) to it.

\subsubsection{CIFAR10 dataset and convergence issues}

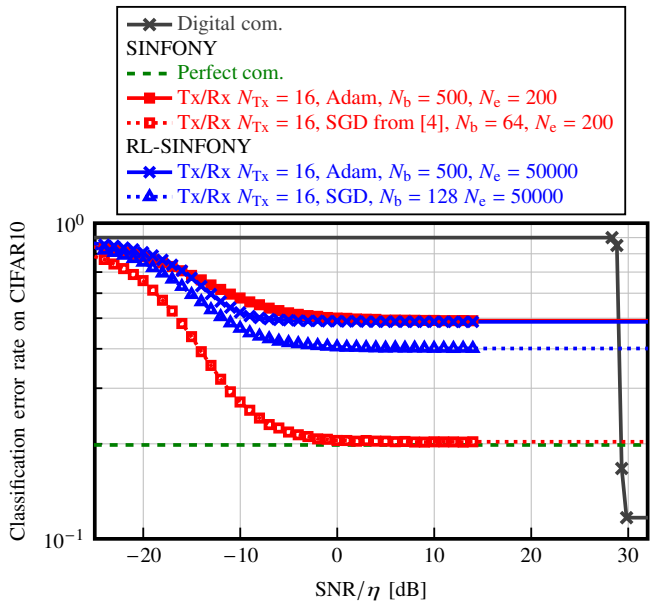
\begin{figure}[!t]
	\centerline{\input{./TikZ/comsem_image_cifar_snr-4_6_RL_normSNR.tikz}}
	\caption{Comparison of the classification error rate of \sinfonirl\ and \sinfoni\ with $\ntx=16$ on CIFAR10 as a function of normalized SNR.}
	\label{fig:sinfony_cifar_RL}
\end{figure}

We further evaluate \sinfonirl\ on the more challenging CIFAR10 validation dataset with $\ntx=16$ and fine-tuned learning rate $\lr=10^{-4}$. The performance curves of \sinfoni\ and \sinfonirl\ with Adam, depicted in Fig.~\ref{fig:sinfony_cifar_RL}, closely align, affirming the effectiveness of \sinfonirl.

Nevertheless, it is crucial to highlight that training with Adam does not converge to a local minimum with the same $80\%$ validation accuracy achieved by the \sinfoni\ benchmark in \cite{beck_semantic_2023}. In that work, we utilized SGD with a batch size of $\Nb=64$, ran for $\Ne=200$ epochs, and employed a dedicated learning rate schedule. Despite exploring various hyperparameter settings, replicating the same performance with \sinfonirl\ has proven elusive.

Additionally, we observed that the training of \sinfonirl\ on the CIFAR10 dataset exhibits slow convergence. For example, using SGD with $\Nb=128$ and $\lr=10^{-4}$ (see Fig.~\ref{fig:sinfony_cifar_RL}), we achieve a validation accuracy of $50\%$ at high SNR after $\Ne+\Nerx=5000+1000=6000$ epochs, still gradually improving to a maximum of $60\%$ after an extensive training period of $\Ne+\Nerx=50000+10000=60000$ epochs.

We assume the slow convergence to be caused by the high variance of the \reinforce\ gradient~\eqref{eq:reinforce2}, which increases by decreasing $\perstd^2$ %
and increasing the continuous output space $\ntx$ of $\xvec$. Training with the more challenging CIFAR10 dataset may require more accurate gradient estimates compared to MNIST. Thus, we suggest exploring variance-reduction techniques in future work~\cite{greensmith_variance_2004, simeone2018brief}. Note that, analogous to the mean $\xvecexpl=\encdet$ of the Gaussian policy \eqref{eq:gaussian_policy}, also the exploration variance $\perstd^2$ can be parametrized by a DNN with shared parameters $\txpars$ or independent parameters $\txpars=[\txpars_{\xvecexpl}, \txpars_{\perstd^2}]^{\trapo}$. Both approaches could facilitate quicker convergence and more efficient hyperparameter tuning, ultimately leading to higher validation accuracy.

\section{Conclusion}
\label{sec:conclusion}

In this work, we expanded on our previous idea from~\cite{beck_semantic_2023} by introducing the Stochastic Policy Gradient (SPG): We designed a semantic communication system via reinforcement learning, separating transmitter and receiver, and not requiring a known or differentiable channel model -- a crucial step towards deployment in practice. Further, we derived the use of the SPG for both classic and semantic communication from the maximization of the mutual information between received and target variables. Numerical results show that our approach achieves comparable performance to a model-aware approach, albeit at the cost of a decreased convergence rate by at least a factor of $10$. It remains the question of how to improve the convergence rate with more challenging datasets.

\bibliographystyle{IEEEtran}
\bibliography{IEEEabrv, references}

\end{document}

%% file: definitions.tex
\newcommand{\aprob}{\ensuremath{q}} 				%
\newcommand{\param}{\ensuremath{\theta}} 			%
\newcommand{\params}{\ensuremath{\bog{\param}}} 	%
\newcommand{\Nsamp}{\ensuremath{N}} 				%

\newcommand{\bo}[1]{\ensuremath{\mathbf{#1}}}
\newcommand{\bog}[1]{\ensuremath{\bm{#1}}}
\newcommand{\txt}[1]{\ensuremath{\textrm{#1}}}

\newcommand{\st}{\ensuremath{\quad \txt{s.t.} \quad}} %
\newcommand{\eqpoint}{\ensuremath{\,.}} %

\DeclareMathOperator*{\armax}{arg\,max}

\newcommand{\argmax}[1][\cdot]{\ensuremath{\underset{#1}{\armax}\ }}

\newcommand{\lpa}{\ensuremath\left(}
\newcommand{\rpa}{\ensuremath\right)}
\newcommand{\lbb}{\ensuremath\left[}
\newcommand{\rbb}{\ensuremath\right]}
\newcommand{\lcb}{\ensuremath\left\{}
\newcommand{\rcb}{\ensuremath\right\}}

\newcommand{\summ}[2][]{\ensuremath{\sum \limits_{#2}^{#1}}} %

\DeclareMathOperator{\E}{E} %
\newcommand{\eval}[2][\cdot]{\ensuremath \E_{#1}\negthinspace\lbb#2\rbb} 	%
\newcommand{\evalnb}[2][\cdot]{\ensuremath \E_{#1}\negthinspace[#2]} 		%
\newcommand{\he}{\mathcal{H}}												%
\newcommand{\HE}[1][\cdot]{\ensuremath \he\lpa#1\rpa} 						%

\newcommand{\rv}{RV}
\newcommand{\Nb}{\ensuremath{N_{\txt{b}}}}                  %
\newcommand{\Ne}{\ensuremath{N_{\txt{e}}}}                  %
\newcommand{\stddev}{\ensuremath{\sigma}}                   %
\newcommand{\noisestd}{\ensuremath{\stddev_{\txt{n}}}}      %
\newcommand{\noisevar}{\ensuremath{n}}                      %
\newcommand{\noisevec}{\ensuremath{\bo{\noisevar}}}         %
\newcommand{\Nclass}{\ensuremath{M}}                        %
\newcommand{\xvar}{\ensuremath{x}}                          %
\newcommand{\xvec}{\ensuremath{\bo{x}}}                     %
\newcommand{\yvar}{\ensuremath{y}}                          %
\newcommand{\yvec}{\ensuremath{\bo{\yvar}}}                 %
\newcommand{\prob}{\ensuremath{p}}                          %
\newcommand{\indi}{\ensuremath{i}}                          %

\newcommand{\semvar}{\ensuremath{s}} 							%
\newcommand{\semvarreal}{\ensuremath{\mathsf{\semvar}}} 		%
\newcommand{\semvec}{\ensuremath{\bo{\semvar}}} 				%
\newcommand{\semvecreal}{\ensuremath{\bog{\semvarreal}}} 		%
\newcommand{\semhvec}{\ensuremath{\hat{\semvec}}} 				%
\newcommand{\rcind}[1][]{\ensuremath \ifthenelse{\equal{#1}{}}{R_{\text{C}}}{R_{\text{C},#1}}} %
\newcommand{\MI}{\ensuremath I} 								%
\newcommand{\mi}[2][]{\ensuremath \MI_{#1}\lpa#2\rpa} 			%
\newcommand{\Lossf}{\ensuremath \mathcal{L}} 					%
\newcommand{\mic}{\ensuremath \MI_{\txt{C}}} 					%
\newcommand{\relvar}{\ensuremath{z}} 							%
\newcommand{\relvarreal}{\ensuremath{\mathsf{z}}} 				%
\newcommand{\relvecest}{\ensuremath{\hat{\bo{\relvar}}}} 		%
\newcommand{\relvec}{\ensuremath{\bo{\relvar}}} 				%
\newcommand{\relvecreal}{\ensuremath{\bog{\relvarreal}}} 		%
\newcommand{\ntx}{\ensuremath{N_{\txt{Tx}}}} 					%
\newcommand{\nrx}{\ensuremath{N_{\txt{w}}}} 					%
\newcommand{\Nrx}{\ensuremath{N_{\txt{Rx}}}} 					%
\newcommand{\Nfeat}{\ensuremath{N_{\txt{Feat}}}} 				%
\newcommand{\speceff}{\ensuremath{\eta}} 						%
\newcommand{\sinfoni}{SINFONY} 									%
\newcommand{\txrx}{Tx/Rx} 					                    %
\newcommand{\perfcomm}{Perfect com.} 		                    %
\newcommand{\sinfoniperfcomm}{\sinfoni\ - \perfcomm} 		    %
\newcommand{\centralimagecomm}{Digital com.} 	                %
\newcommand{\Nsemvec}{\ensuremath{N_{\semvar}}} 				%
\newcommand{\Nrelvec}{\ensuremath{N_{\relvar}}} 				%
\newcommand{\Nimx}{\ensuremath{N_{\txt{x}}}} 					%
\newcommand{\Nimy}{\ensuremath{N_{\txt{y}}}} 					%
\newcommand{\Nimc}{\ensuremath{N_{\txt{c}}}} 					%
\newcommand{\setsym}{\ensuremath{M}} 							%
\newcommand{\dom}{\ensuremath{\mathcal{\setsym}}} 				%
\newcommand{\semset}{\ensuremath{\dom_{\semvar}}} 				%
\newcommand{\relset}{\ensuremath{\dom_{\relvar}}} 				%
\newcommand{\xset}{\ensuremath{\dom_{\xvar}}} 					%
\newcommand{\yset}{\ensuremath{\dom_{\yvar}}} 					%
\newcommand{\txpars}{\ensuremath{\params}} 						%
\newcommand{\Ntxpars}{\ensuremath{N_{\txpars}}} 				%
\newcommand{\Nrxpars}{\ensuremath{N_{\rxpars}}} 				%
\newcommand{\rxpar}{\ensuremath{\varphi}} 						%
\newcommand{\rxpars}{\ensuremath{\bog{\rxpar}}} 				%
\newcommand{\covariance}{\ensuremath{\bog{\Sigma}}} 			%
\newcommand{\mean}{\ensuremath{\mu}} 							%
\newcommand{\funcrtrick}{\ensuremath{f}} 						%
\newcommand{\meanfunc}{\ensuremath{\mean}} 						%
\newcommand{\dirac}{\ensuremath{\delta}} 						%
\newcommand{\ceparams}{\ensuremath{\Lossf_{\txpars, \rxpars}^{\txt{CE}}}}	%

\newcommand{\perstd}{\ensuremath{\stddev_{\txt{exp}}}} 			        %
\newcommand{\spgobj}{\ensuremath{\Lossf_{\txpars}^{\txt{SPG}}}}	        %
\newcommand{\sinfonirl}{RL-\sinfoni}							        %
\newcommand{\xvecexpl}{\ensuremath{\bar{\xvec}}} 		                %
\newcommand{\encdet}{\ensuremath{\meanfunc_{\txpars}(\semvec)}} 		%
\newcommand{\Nerx}{\ensuremath{N_{\txt{e,rx}}}}                         %
\newcommand{\lr}{\ensuremath{\epsilon}}                                 %
\newcommand{\xvarreal}{\ensuremath{\mathsf{\xvar}}} 		            %
\newcommand{\xvecreal}{\ensuremath{\bog{\xvarreal}}} 		            %
\newcommand{\yvarreal}{\ensuremath{\mathsf{\yvar}}} 		            %
\newcommand{\yvecreal}{\ensuremath{\bog{\yvarreal}}} 		            %
\newcommand{\noisevarreal}{\ensuremath{\mathsf{\noisevar}}}             %
\newcommand{\noisevecreal}{\ensuremath{\bog{\noisevarreal}}}            %
\newcommand{\xvecexplreal}{\ensuremath{\bar{\bog{\mathsf{\xvar}}}}}     %
\newcommand{\snr}{\ensuremath{\txt{SNR}}}                               %
\newcommand{\reinforce}{REINFORCE} 		                                %
\newcommand{\dataset}[1][]{\ensuremath\mathcal{D}_{#1}}                 %
\newcommand{\trainingset}{\ensuremath\dataset[\text{T}]}                %
\newcommand{\pilotset}{\ensuremath\dataset[\text{P}]}                   %
\newcommand{\semtrain}{\ensuremath{\semvecreal_{\indi}}} 	            %
\newcommand{\ytrain}{\ensuremath{\yvecreal_{\indi}}} 		            %
\newcommand{\xtrain}{\ensuremath{\xvecreal_{\indi}}} 		            %
\newcommand{\reltrain}{\ensuremath{\relvecreal_{\indi}}} 	            %
\newcommand{\xexpltrain}{\ensuremath{\xvecexplreal_{\indi}}} 	        %
\newcommand{\Npilot}{\ensuremath{N_{\txt{pilot}}}}                      %

\newcommand{\Ntrain}{\ensuremath{N_{\txt{train}}}}                  %

\newcommand{\D}{\ensuremath{\mathrm{d}}} %

\newcommand{\lnorm}[1][\cdot]{\ensuremath{l_{#1}}}

\newcommand{\trapo}{\ensuremath{T}} %
\newcommand{\realnum}{\ensuremath{\mathbb{R}}} %
\newcommand{\eye}{\ensuremath{\bo{I}}} %
\newcommand{\normdis}{\ensuremath{\mathcal{N}}} %

\newcommand{\idxC}{\ensuremath k} %

\newcommand{\mF}[1][\nC]{\ensuremath \mathbf{F}_{\nC}} %

\newcommand{\eA}{\ensuremath a} %
\newcommand{\setA}[1][]{\ensuremath\mathcal{\MakeUppercase{\eA}}_{#1}} %
\newcommand{\defD}[1][0]{\ensuremath\lcb\vdwir[\idxC]\,\vert\,\sd_{\idxC,\idxA}\in\ifthenelse{#1>0}{\setA[\idxC,\idxA]}{\setA}\quad\forall\,\idxA=1,\dotsc,2\nA\rcb} %

\newcommand{\nA}{\ensuremath N_{\text{L}}} %
\newcommand{\nC}{\ensuremath N_{\text{C}}} %

\newcommand{\vrc}[1][]{\ensuremath \ifthenelse{\equal{#1}{}}{\mathbf{r}_{\text{C}}}{\mathbf{r}_{\text{C},#1}}} %
\newcommand{\rce}[1][]{\ensuremath \ifthenelse{\equal{#1}{}}{\hat{R}_{\text{C}}}{\hat{R}_{\text{C},#1}}} %

\newcommand{\drc}[1][]{\ensuremath \ifthenelse{\equal{#1}{}}{R'_{\text{C}}}{R'_{\text{C},#1}}} %

\newcommand{\sd}{\ensuremath d} %

%% file: plotopt.tex
\usepackage{pgfplots}
\pgfplotsset{compat=newest}
\usetikzlibrary{pgfplots.groupplots}
\usetikzlibrary{calc,shapes.geometric,shapes.misc,arrows,matrix}
\usepgfplotslibrary{fillbetween}
\usepackage[absolute,overlay]{textpos} %

\definecolor{darkgreen}{rgb}{0,0.498039,0}
\definecolor{darkyellow}{rgb}{0.75,0.75,0}
\definecolor{magenta}{rgb}{0.75,0,0.75}
\newcommand{\clsd}{darkgray} %
\newcommand{\clmmse}{darkgray} %
\newcommand{\clmosic}{purple} %
\newcommand{\clamp}{darkgreen} %
\newcommand{\clsdr}{orange} %
\newcommand{\clmmnet}{darkyellow} %
\newcommand{\cldetnet}{magenta} %
\newcommand{\cloamp}{blue} %
\newcommand{\clcmd}{red} %
\newcommand{\clcmdshal}{orange} %
\newcommand{\clhcmd}{darkgreen} %
\newcommand{\clawgn}{green!50!black} %

\pgfplotsset{
	ptsd/.style={color=\clsd, mark=*, mark size=1.4
		}, %
	ptmmse/.style={color=\clmmse, mark=triangle*, dashed, mark options={solid}, mark size=1.4
		}, %
	ptmosic/.style={color=\clmosic, mark=triangle*, mark size=1.4
		}, %
	ptamp/.style={color=\clamp, mark=diamond*, mark size=1.4
		}, %
	ptsdr/.style={color=\clsdr, mark=*, dashed, mark options={solid}, mark size=1.4
		}, %
	ptdetnet/.style={color=\cldetnet, mark=square*, dashed, mark options={solid}, mark size=1.4
		}, %
	ptmmnet/.style={color=\clmmnet, mark=diamond*, dashed, mark options={solid}, mark size=1.4
		}, %
	ptoamp/.style={color = \cloamp, mark=x, mark size=4*0.7, mark options={solid}
		}, %
	ptcmd/.style={color=\clcmd, mark=square*, mark size=1.4
		}, %
	ptcmdshal/.style={\clcmdshal, dashed, mark=square*, mark options={solid}, mark size=2*0.7
		}, %
	pthcmd/.style={color=\clhcmd, mark size=1.4, mark options={solid}, %
		}, %
	ptawgn/.style={color=\clawgn, semithick
		}, %
}

\pgfplotsset{
	every axis plot/.style={line width=1.4pt},
	tick pos=left,
	yminorgrids,
	every outer x axis line/.style={line width=1.4pt},
	every outer y axis line/.style={line width=1.4pt},
	every tick/.style={
			black,
			line width=0.7pt,
		},
	label style={font=\fontsize{8}{9}\selectfont},
	title style={font=\fontsize{8}{9}\selectfont},
	every y tick label/.style={font=\fontsize{8}{9}\color{black}},
	every x tick label/.style={font=\fontsize{8}{9}\color{black}},
	every extra x tick/.style={grid style={solid, violet, line width=2.8pt},
			x tick label style={/pgf/number format/.cd,precision=10}
		},
	legend style= {
			line width=0.7pt,
			font=\fontsize{11.4}{12}\selectfont\color{black},
			legend cell align=left,
			align=left,
			fill=white,
			fill opacity=0.8, draw opacity=1, text opacity=1,
			nodes={scale=0.7, transform shape},
		},
	height = 8cm, %
}

%% file: TikZ/com_system_model_basic_small_RL_v2.tikz
\begin{tikzpicture}[%
					inner xsep = 0pt, inner ysep = 0pt, %
					outer xsep = 0pt, outer ysep = 0pt, %
					> = latex,
					ucircle/.style = {circle, line width = 0.30mm, minimum size = 4.00mm, draw = black, fill = white},
					urec/.style = {rectangle, line width = 0.60mm, draw = black, fill = white, minimum width = 50.0, minimum height = 20.00},
					invrec/.style = {rectangle, line width = 0.30mm, draw = none, fill = none, minimum width = 50.0, minimum height = 20.00},				
					arrowthick/.style = {->, line width = 0.6mm},
					arrowthin/.style = {<->, line width = 0.3mm},
					font = {\fontsize{7pt}{12}\selectfont},
					linesplit/.style = {circle, draw = black, fill = black, inner sep = 0pt, minimum height = 2},
					linethick/.style = {-, line width = 0.60mm},
					linethin/.style={-,line width = 0.30mm},
					]

\newcommand{\clsem}{blue} %
\newcommand{\clenc}{darkgreen} %
\newcommand{\clchan}{gray} %
\newcommand{\clencchan}{orange} %
\newcommand{\cldec}{red} %

\node[urec, minimum height=40.00, minimum width=40.0, text width=2cm, align=center, draw = \clsem, rounded corners=5] (App1) at (-14,0) {Semantic Source\\ $\relvec\sim \prob(\relvec)$};
\node[urec, minimum height=40.00, minimum width=40.0, text width=2cm, align=center, draw = \clsem, rounded corners=5] (SChan) at (-11.25,0) {Semantic Channel\\ $\prob(\semvec|\relvec)$};
\node[urec, minimum height=30.00, minimum width=60.0, text width=2.3cm, align=center, draw = \clenc, rounded corners=5] (Enc) at (-8.0,0) {Encoder\\ $\prob_{\txpars}(\xvec|\semvec)$};
\node[urec, minimum height=30.00, minimum width=60.0, text width=2.3cm, align=center, draw = \clchan, rounded corners=5] (Chan) at (-8.0,-1.75) {Communication Channel $\prob(\yvec|\xvec)$};
\node[urec, minimum height=30.00, minimum width=60.0, text width=2.3cm, align=center, draw = \cldec, rounded corners=5] (Dec) at (-11.25,-1.75) {Semantic Decoder\\ $\aprob_{\rxpars}(\relvec|\yvec)$};
\node[urec, minimum height=40.00, minimum width=40.0, text width=2cm, align=center, draw = \clsem, rounded corners=5] (App1_rx) at (-14,-1.75) {Semantic Estimate $\relvecest$};

\node[urec, text depth = 3 cm, minimum height=95.00, minimum width=92.5, fill=none, dashed, dash pattern={on 7pt off 3pt}, rounded corners=10, draw = \clencchan] (Enc2) at (-8.2,-0.75) {};
\node[anchor = north east, inner xsep = 0.00cm, inner ysep = 0.08cm, xshift = 0.0cm, text width=2cm, align=center] at (Enc2.north) {\color{\clencchan}{$\prob_{\txpars}(\yvec|\semvec)$}};

\draw[arrowthick] (App1) -> (SChan) node[midway, above, rotate=0, align=center, anchor=south, yshift = 0.2cm] {$\relvec$};
\draw[arrowthick] (SChan) -> (Enc) node[midway, above, rotate=0, align=center, anchor=south, yshift = 0.2cm] {$\semvec$};
\draw[arrowthick] (Enc) -> (Chan) node[midway, above, rotate=0, align=center, anchor=west, yshift = 0.0cm, xshift=0.2cm] {$\xvec$};
\draw[arrowthick] (Chan) -> (Dec)  node[midway, above, rotate=0, align=center, anchor=west, yshift = 0.2cm] {$\yvec$};
\draw[arrowthick] (Dec) -> (App1_rx) node[pos=0.5, above, rotate=0, yshift = 0.2cm] {$\relvec$};

\end{tikzpicture}%

%% file: TikZ/com_system_model_basic_RL.tikz
\begin{tikzpicture}[%
		inner xsep = 0pt, inner ysep = 0pt, %
		outer xsep = 0pt, outer ysep = 0pt, %
		> = latex,
		ucircle/.style = {circle, line width = 0.30mm, minimum size = 4.00mm, draw = black, fill = white},
		urec/.style = {rectangle, line width = 0.60mm, draw = black, fill = white, minimum width = 50.0, minimum height = 20.00},
		invrec/.style = {rectangle, line width = 0.30mm, draw = none, fill = none, minimum width = 50.0, minimum height = 20.00},
		arrowthick/.style = {->, line width = 0.6mm},
		arrowthin/.style = {<->, line width = 0.3mm},
		font = {\fontsize{7pt}{12}\selectfont},
		linesplit/.style = {circle, draw = black, fill = black, inner sep = 0pt, minimum height = 2},
		linethick/.style = {-, line width = 0.60mm},
		linethin/.style={-,line width = 0.30mm},
	]

	\newcommand{\clsem}{blue} %
	\newcommand{\clenc}{darkgreen} %
	\newcommand{\clchan}{gray} %
	\newcommand{\cldec}{red} %
	\newcommand{\clopt}{black} %

	\node[urec, minimum height=30.00, minimum width=85.0, text width=3cm, align=center, draw = \clenc, rounded corners=5] (Enc) at (-8.0,0) {Encoder\\ $\prob_{\txpars}(\xvecexpl|\semvec)=\dirac(\xvecexpl-\encdet)$};
	\node[urec, minimum height=30.00, minimum width=60.0, text width=2.2cm, align=center, draw = \clenc, rounded corners=5] (Sample) at (-8,-1.5) {Sampler / Explorer\\ $\prob(\xvec|\xvecexpl)$};
	\node[urec, minimum height=30.00, minimum width=60.0, text width=2.2cm, align=center, draw = \clchan, rounded corners=5] (Chan) at (-8,-3) {Communication Channel $\prob(\yvec|\xvec)$};
	\node[urec, minimum height=30.00, minimum width=85.0, text width=3cm, align=center, draw = \cldec, rounded corners=5] (Dec) at (-8,-4.5) {Semantic Decoder\\ $\aprob_{\rxpars}(\relvec|\yvec)$};
	\node[urec, minimum height=30.00, minimum width=30.0, text width=2.2cm, align=center, draw = \clenc, rounded corners=0] (Txopt) at (-5,0) {2. Train Encoder\\ $\frac{\partial}{\partial\txpars}\spgobj$};
	\node[urec, minimum height=30.00, minimum width=30.0, text width=2.2cm, align=center, draw = \cldec, rounded corners=0] (Rxopt) at (-5,-5.5) {1. Train Decoder\\ $\frac{\partial}{\partial\rxpars}\ceparams$};
	\node[linesplit, minimum height=5] (hnode) at (-5,-4.5) {};

	\draw[arrowthick] (-10.5,0) -> (Enc) node[midway, above, rotate=0, align=center, anchor=south, yshift = 0.1cm] {$\semtrain\in\trainingset$};
	\draw[arrowthick] (Enc) -> (Sample) node[midway, above, rotate=0, align=center, anchor=west, yshift = 0.0cm, xshift=0.2cm] {$\xexpltrain$};
	\draw[arrowthick] (Sample) -> (Chan) node[midway, above, rotate=0, align=center, anchor=west, yshift = 0.0cm, xshift=0.2cm] {$\xtrain\in\trainingset$};
	\draw[arrowthick] (Chan) -> (Dec)  node[midway, above, rotate=0, align=center, anchor=west, xshift = 0.2cm] {$\ytrain\in\trainingset$};
	\draw[arrowthick] (Dec) -> (-10.5,-4.5) node[pos=0.5, above, rotate=0, yshift = 0.1cm] {$\reltrain\in\trainingset$};
	\draw[arrowthick, \clenc] (hnode) -> (Txopt) node[pos=0.5, xshift = 0.1cm, anchor=west, align=center, text width=2cm, color = \clenc] {Feedback link:\\ $-\ln \aprob_{\rxpars}(\reltrain|\ytrain)$};
	\draw[arrowthick, \cldec] (hnode) -> (Rxopt) node[pos=0.75, xshift = 0.1cm, anchor=west, align=center, text width=2cm] {};
	\draw[linethick] (Dec) -- (hnode) node[] {};
	\draw[arrowthick, \clenc] (Txopt) -> (Enc) node[pos=0.5, align=center, text width=2cm, xshift=0.3cm, yshift=-0.95cm] {Reinforcement Learning};
	\draw[arrowthick, \cldec] (Rxopt) -| (Dec) node[pos=0.25, xshift = 0.0cm, anchor=center, align=center, text width=1.25cm] {Supervised Learning};

\end{tikzpicture}%

%% file: TikZ/distributed_resnet_v5_small_RL.tikz
\begin{tikzpicture}[%
					inner xsep=0pt, inner ysep=0pt, %
					outer xsep=0pt, outer ysep=0pt, %
					>=latex,
					ucircle/.style={circle,line width=0.30mm,minimum size=4.00mm,draw=black,fill=white},
					urec/.style={rectangle,line width=0.60mm,draw=black,fill=white,minimum width=50.0,minimum height=20.00},
					invrec/.style={rectangle,line width=0.30mm,draw=none,fill=none,minimum width=50.0,minimum height=20.00},
					arrowthick/.style={->,line width=0.6mm},
					arrowthin/.style={<->,line width=0.6mm},
					linesplit/.style={circle, draw=black, fill=black, inner sep=0pt, minimum height=2},
					font={\fontsize{7pt}{12}\selectfont},
					]

\newcommand{\clsem}{blue} %
\newcommand{\clsemchan}{blue!70} %
\newcommand{\clenc}{darkgreen} %
\newcommand{\clresnet}{\clamp} %
\newcommand{\clchan}{gray} %
\newcommand{\cltrans}{purple} %
\newcommand{\cldec}{red} %

\node[urec, draw = \clsem, minimum height=10.00, minimum width=30.0, align=center, rounded corners=5] (z0) at (-10.125,1.25) {$\relvec$};

\node[linesplit, minimum height=5] (rvnode1) at (-11.25,1.25) {};
\node[linesplit, minimum height=5] (rvnode2) at (-9,1.25) {};

\node[urec, draw = \clsemchan, minimum height=20.00, minimum width=40.0, text width=1.3cm, align=center, rounded corners=5] (image1) at (-13.5,0.5) {$\prob(\semvec_1|\relvec)$};
\node[urec, draw = \clenc, minimum height=30.00, minimum width=50.0, text width=1.8cm, fill=none, rounded corners=5, align=center] (Enc1) at (-13.5,-1) {Encoder 1: $\prob_{\txpars_1}(\xvec_1|\semvec_1)$};
\node[urec, draw = \clchan, minimum height=20.00, minimum width=40.0, text width=1.3cm, align=center, rounded corners=5] (ch1) at (-13.5,-2.5) {Channel 1};
\draw[arrowthick] (image1) -> (Enc1) node[pos=0.35, rotate=0, align=center, anchor=west, xshift = 0.1cm] {$\semvec_1 \in \realnum^{\Nimx\times \Nimy\times \Nimc}$};
\draw[arrowthick] (Enc1) -> (ch1) node[midway, rotate=0, align=center, anchor=west, xshift = 0.1cm, yshift = 0.15cm] {$\xvec_1 \in \realnum^{\ntx\times 1}$};

\node[urec, draw = \clsemchan, draw = teal, minimum height=20.00, minimum width=40.0, text width=1.3cm, align=center, rounded corners=5] (image2) at (-11.25,0.5) {$\prob(\semvec_2|\relvec)$};
\node[urec, draw = \clenc, minimum height=30.00, minimum width=50.0, text width=1.8cm, fill=none, rounded corners=5, align=center] (Enc2) at (-11.25,-1) {Encoder 2: $\prob_{\txpars_2}(\xvec_2|\semvec_2)$};
\node[urec, draw = \clchan, minimum height=20.00, minimum width=40.0, text width=1.3cm, align=center, rounded corners=5] (ch2) at (-11.25,-2.5) {Channel 2};
\draw[arrowthick] (image2) -> (Enc2) node[pos=0.35, rotate=0, align=center, anchor=west, xshift = 0.1cm] {$\semvec_2$};
\draw[arrowthick] (Enc2) -> (ch2) node[midway, rotate=0, align=center, anchor=west, xshift = 0.1cm, yshift = 0.1cm] {$\xvec_2$};%

\node[urec, draw = \clsemchan, draw = cyan, minimum height=20.00, minimum width=40.0, text width=1.3cm, align=center, rounded corners=5] (image3) at (-9,0.5) {$\prob(\semvec_3|\relvec)$};
\node[urec, draw = \clenc, minimum height=30.00, minimum width=50.0, text width=1.8cm, fill=none, rounded corners=5, align=center] (Enc3) at (-9,-1) {Encoder 3: $\prob_{\txpars_3}(\xvec_3|\semvec_3)$};
\node[urec, draw = \clchan, minimum height=20.00, minimum width=40.0, text width=1.3cm, align=center, rounded corners=5] (ch3) at (-9,-2.5) {Channel 3};
\draw[arrowthick] (image3) -> (Enc3) node[pos=0.35, rotate=0, align=center, anchor=west, xshift = 0.1cm] {$\semvec_3$};
\draw[arrowthick] (Enc3) -> (ch3)  node[midway, rotate=0, align=center, anchor=west, xshift = 0.1cm, yshift = 0.1cm] {$\xvec_3$};%

\node[urec, draw = \clsemchan, draw = lightgray, minimum height=20.00, minimum width=40.0, text width=1.3cm, align=center, rounded corners=5] (image4) at (-6.75,0.5) {$\prob(\semvec_4|\relvec)$};
\node[urec, draw = \clenc, minimum height=30.00, minimum width=50.0, text width=1.8cm, fill=none, rounded corners=5, align=center] (Enc4) at (-6.75,-1) {Encoder 4: $\prob_{\txpars_4}(\xvec_4|\semvec_4)$};
\node[urec, draw = \clchan, minimum height=20.00, minimum width=40.0, text width=1.3cm, align=center, rounded corners=5] (ch4) at (-6.75,-2.5) {Channel 4};
\draw[arrowthick] (image4) -> (Enc4) node[pos=0.35, rotate=0, align=center, anchor=west, xshift = 0.1cm] {$\semvec_4$};
\draw[arrowthick] (Enc4) -> (ch4) node[midway, rotate=0, align=center, anchor=west, xshift = 0.1cm, yshift = 0.1cm] {$\xvec_4$};%

\draw[arrowthick] (z0) -| (image1);
\draw[arrowthick,-] (z0) -> (rvnode1);
\draw[arrowthick,-] (z0) -> (rvnode2);
\draw[arrowthick] (z0) -| (image4);
\draw[arrowthick] (rvnode1) -> (image2);
\draw[arrowthick] (rvnode2) -> (image3);

\node[urec, draw = \clsem, minimum height=10.00, minimum width=30.0, align=center, rounded corners=5] (z) at (-8.0,-4.15) {$\relvecest$};
\node[urec, draw = \cldec, minimum height=30.00, minimum width=50.0, text width=1.8cm, fill=none, rounded corners=5, align = center] (Dec) at (-10.125,-4.15) {Decoder:\\ $\aprob_{\rxpars}(\relvec|\yvec)$};

\draw[arrowthick] (ch1.south) -- (-13.5,-3.25) node[midway, rotate=0, align=center, anchor=west, yshift = 0.05cm, xshift = 0.1cm] {$\yvec_1 \in \realnum^{\Nrx\times 1}$} -| ([xshift=-0.45cm]Dec.north);
\draw[arrowthick] (ch2.south) -- (-11.25,-3.15) node[midway, rotate=0, align=center, anchor=east, yshift = -0.025cm, xshift = -0.1cm] {$\yvec_2$} -| ([xshift=-0.15cm]Dec.north);
\draw[arrowthick] (ch3.south) -- (-9,-3.15) node[midway, rotate=0, align=center, anchor=west, yshift = -0.025cm, xshift = 0.1cm] {$\yvec_3$} -| ([xshift=0.15cm]Dec.north);
\draw[arrowthick] (ch4.south) -- (-6.75,-3.25) node[midway, rotate=0, align=center, anchor=west, yshift = 0.0cm, xshift = 0.1cm] {$\yvec_4$} -| ([xshift=0.45cm]Dec.north);

\draw[arrowthick] (Dec) -> (z);

\end{tikzpicture}%

%% file: TikZ/comsem_image_mnist_snr-4_6_RL_normSNR.tikz
\begin{tikzpicture}%
\begin{axis}[
width=10.5cm,
height=6cm, %
scale=0.7,
scale only axis,
xlabel={$\snr/\speceff$ [dB]},
xmajorgrids,
xmin=-17, xmax=21,
xminorgrids,
ylabel={Classification error rate on MNIST},
ymajorgrids,
ymin=3.5e-3, ymax=1, %
yminorgrids,
ymode=log,
axis background/.style={fill=white},
legend style={at={(0.30, 0.65)}, anchor=west},
legend entries={{\color{\clsd} \centralimagecomm},
				{\hspace{-1.075cm} \sinfoni},
				{\color{\clamp} \perfcomm},
				{\color{\clcmd} Tx/Rx $\ntx=14$, Adam $\Ne=100$},
				{\hspace{-1.075cm} \sinfonirl},
				{\color{\cloamp} Tx/Rx $\ntx=14$, Adam $\Ne=3000$}, %
				},
]
\addplot [ptsd, mark=x, mark size=3, mark options={solid}]
table {%
-20 0.90
17.5258751031689 0.90204
18.0258751031689 0.8696
18.5258751031689 0.00885999999999998
19.0258751031689 0.00400000810623169
25 0.00400000810623169
};
\addplot [ptoamp, empty legend, dashed, mark = None, mark options={solid}]
table {%
0 0
1 2
};
\addplot[ptamp, dashed, mark = None, mark options={solid}]
table {%
-20 0.008899986743927
22 0.008899986743927
};
\addplot[ptcmd]
table {%
-36.0205999132796 0.872769999504089
-35.0205999132796 0.867739999294281
-34.0205999132796 0.865269999206066
-33.0205999132796 0.85799999833107
-32.0205999132796 0.852420000731945
-31.0205999132796 0.845870000123978
-30.0205999132796 0.836549998819828
-29.0205999132796 0.829839998483658
-28.0205999132796 0.817860002815723
-27.0205999132796 0.805890001356602
-26.0205999132796 0.79108000099659
-25.0205999132796 0.771200002729893
-24.0205999132796 0.749849998950958
-23.0205999132796 0.725510001182556
-22.0205999132796 0.692640000581741
-21.0205999132796 0.660130003094673
-20.0205999132796 0.619489997625351
-19.0205999132796 0.577829995751381
-18.0205999132796 0.523819994926453
-17.0205999132796 0.46269000172615
-16.0205999132796 0.399520003795624
-15.0205999132796 0.337509995698929
-14.0205999132796 0.270060002803802
-13.0205999132796 0.206629997491837
-12.0205999132796 0.152589994668961
-11.0205999132796 0.108989995718002
-10.0205999132796 0.0755299985408782
-9.02059991327963 0.0545599937438965
-8.02059991327963 0.0415000021457672
-7.02059991327962 0.0328099966049196
-6.02059991327962 0.0271400034427643
-5.02059991327962 0.0245200037956238
-4.02059991327962 0.0215899944305421
-3.02059991327962 0.0200399994850158
-2.02059991327962 0.0187099874019622
-1.02059991327962 0.0181599974632263
-0.0205999132796242 0.0177599906921387
0.979400086720376 0.0171600043773651
1.97940008672038 0.0166299998760223
2.97940008672038 0.0164400041103362
3.97940008672038 0.0159100055694581
4.97940008672038 0.0159000039100647
5.97940008672038 0.0157900035381318
6.97940008672038 0.0160600006580353
7.97940008672038 0.0160199999809265
8.97940008672037 0.0157499969005584
9.97940008672037 0.0157000005245209
10.9794000867204 0.0157599925994872
11.9794000867204 0.0156499981880189
12.9794000867204 0.0156099975109101
13.9794000867204 0.0157299995422362
22 0.0157299995422362
};
\addplot [ptoamp, empty legend, dashed, mark = None, mark options={solid}]
table {%
0 0
1 2
};
\addplot[ptoamp]
table {%
-36.0205999132796 0.892289996147156
-35.0205999132796 0.892520010471344
-34.0205999132796 0.890410006046295
-33.0205999132796 0.889320015907288
-32.0205999132796 0.884190022945404
-31.0205999132796 0.882109999656677
-30.0205999132796 0.877680003643036
-29.0205999132796 0.874210000038147
-28.0205999132796 0.868889987468719
-27.0205999132796 0.861739993095398
-26.0205999132796 0.85139000415802
-25.0205999132796 0.841580033302307
-24.0205999132796 0.828159987926483
-23.0205999132796 0.810890018939972
-22.0205999132796 0.790629982948303
-21.0205999132796 0.765009999275208
-20.0205999132796 0.728909969329834
-19.0205999132796 0.687819957733154
-18.0205999132796 0.635299980640411
-17.0205999132796 0.574040055274963
-16.0205999132796 0.501230001449585
-15.0205999132796 0.420839965343475
-14.0205999132796 0.338380038738251
-13.0205999132796 0.254630029201508
-12.0205999132796 0.179459929466248
-11.0205999132796 0.119480013847351
-10.0205999132796 0.0797699689865112
-9.02059991327963 0.0525899529457092
-8.02059991327963 0.0382900238037109
-7.02059991327962 0.0293000340461731
-6.02059991327962 0.0259000658988953
-5.02059991327962 0.0228000283241272
-4.02059991327962 0.021340012550354
-3.02059991327962 0.0199400186538696
-2.02059991327962 0.0193699598312378
-1.02059991327962 0.0185499787330627
-0.0205999132796242 0.0183500051498413
0.979400086720376 0.0175999402999878
1.97940008672038 0.0177999138832092
2.97940008672038 0.0174499750137329
3.97940008672038 0.0173999071121216
4.97940008672038 0.0173299908638
5.97940008672038 0.0173699855804443
6.97940008672038 0.0172399282455444
7.97940008672038 0.0171299576759338
8.97940008672037 0.0169999599456787
9.97940008672037 0.0171099901199341
10.9794000867204 0.0169699788093567
11.9794000867204 0.0169899463653564
12.9794000867204 0.0170398950576782
13.9794000867204 0.0168900489807129
22 0.0168900489807129
};
\end{axis}

\end{tikzpicture}

%% file: TikZ/comsem_image_mnist_snr-4_6_RL_convergence.tikz
\begin{tikzpicture}%
\begin{axis}[
width=10.5cm,
height=6cm, %
scale=0.7,
scale only axis,
xlabel={Epoch $\Ne$},
xmajorgrids,
xmin=1,
xmax=4e3,
xminorgrids,
xmode=log,
ylabel={Cross-entropy loss on MNIST},
ymajorgrids,
ymin=6.5e-2,
ymax=3,
yminorgrids,
ymode=log,
axis background/.style={fill=white},
legend entries={{\color{\clcmd}\sinfoni\\ \color{\clcmd}Adam $\Ne=100$},
				{\color{\cloamp}\sinfonirl\\ \color{\cloamp}Adam $\Ne=3000$, $\Nerx=600$}, %
				},
]
\addplot[name path = A1, ptcmd, mark = None, line width = 0.7pt, opacity = 0.3, forget plot]
table[x index = 0, y index = 2]{\main/TikZ/data/training_convergence/MNIST/SINFONY_MNIST4_adam.dat};
\addplot[name path = A2, ptcmd, mark = None, line width = 0.7pt, opacity = 0.3, forget plot]
table[x index = 0, y index = 3]{\main/TikZ/data/training_convergence/MNIST/SINFONY_MNIST4_adam.dat};
\addplot[ptcmd, opacity=0.3, forget plot]
fill between[of=A1 and A2];
\addplot[ptcmd, mark = None, line width = 0.7pt, opacity = 1] %
table[x index = 0, y index = 1]{\main/TikZ/data/training_convergence/MNIST/SINFONY_MNIST4_adam.dat};

\addplot[name path = C1, ptoamp, mark = None, line width = 0.7pt, opacity = 0.3, forget plot]
table[x index = 0, y index = 2]{\main/TikZ/data/training_convergence/MNIST/RL-SINFONY_MNIST4_adam.dat};
\addplot[name path = C2, ptoamp, mark = None, line width = 0.7pt, opacity = 0.3, forget plot]
table[x index = 0, y index = 3]{\main/TikZ/data/training_convergence/MNIST/RL-SINFONY_MNIST4_adam.dat};
\addplot[ptoamp, opacity=0.3, forget plot]
fill between[of=C1 and C2];
\addplot[ptoamp, mark = None, line width = 0.7pt, opacity = 1] %
table[x index = 0, y index = 1]{\main/TikZ/data/training_convergence/MNIST/RL-SINFONY_MNIST4_adam.dat};

\end{axis}

\end{tikzpicture}

%% file: TikZ/comsem_image_cifar_snr-4_6_RL_normSNR.tikz
\begin{tikzpicture}%
\begin{axis}[
width=10.5cm,
height=6cm, %
scale=0.7,
scale only axis,
xlabel={$\snr/\speceff$ [dB]},
xmajorgrids,
xmin=-25, xmax=32,
xminorgrids,
ylabel={Classification error rate on CIFAR10},
ymajorgrids,
ymin=1e-1, ymax=1,
yminorgrids,
ymode=log,
axis background/.style={fill=white},
legend style={at={(0.5,1.025)},anchor=south},
legend entries={{\color{\clsd} \centralimagecomm},
				{\hspace{-1.075cm} \sinfoni},
				{\color{\clamp} \perfcomm},
				{\color{\clcmd} Tx/Rx $\ntx=16$, Adam, $\Nb=500$, $\Ne=200$},
				{\color{\clcmd} Tx/Rx $\ntx=16$, SGD from \cite{beck_semantic_2023}, $\Nb=64$, $\Ne=200$},
				{\hspace{-1.075cm} \sinfonirl},
				{\color{\cloamp} Tx/Rx $\ntx=16$, Adam, $\Nb=500$, $\Ne=50000$}, %
				{\color{\cloamp} Tx/Rx $\ntx=16$, SGD, $\Nb=128$ $\Ne=50000$}, %
				},
]
\addplot [ptsd, mark=x, mark size=3, mark options={solid}]
table {%
-30 0.90
28.3294643247982 0.90006
28.8294643247982 0.85027
29.3294643247982 0.1671
29.8294643247982 0.1167
33.8294643247982 0.1167
};
\addplot [ptoamp, empty legend, dashed, mark = None, mark options={solid}]
table {%
0 0
1 2
};
\addplot [ptamp, dashed, mark = None, mark options={solid}]
table {%
-30 0.198400020599365
33 0.198400020599365
};
\addplot[ptcmd, mark options={solid}, opacity=1] %
table {%
-36.0205999132796 0.887370000034571
-35.0205999132796 0.884979999065399
-34.0205999132796 0.882139999419451
-33.0205999132796 0.880280001461506
-32.0205999132796 0.876789999008179
-31.0205999132796 0.872890000045299
-30.0205999132796 0.870259998738766
-29.0205999132796 0.866429999470711
-28.0205999132796 0.859899999201298
-27.0205999132796 0.855129998922348
-26.0205999132796 0.848369999229908
-25.0205999132796 0.84033999890089
-24.0205999132796 0.831470002233982
-23.0205999132796 0.82133000344038
-22.0205999132796 0.811340001225472
-21.0205999132796 0.796649999916553
-20.0205999132796 0.783379998803139
-19.0205999132796 0.766229997575283
-18.0205999132796 0.748749999701977
-17.0205999132796 0.729170003533363
-16.0205999132796 0.706810003519058
-15.0205999132796 0.683219999074936
-14.0205999132796 0.661370000243187
-13.0205999132796 0.637280005216599
-12.0205999132796 0.616889995336533
-11.0205999132796 0.596060001850128
-10.0205999132796 0.579319998621941
-9.02059991327963 0.563109996914864
-8.02059991327963 0.551589998602867
-7.02059991327962 0.53806000649929
-6.02059991327962 0.528559997677803
-5.02059991327962 0.520030003786087
-4.02059991327962 0.514359998703003
-3.02059991327962 0.509650000929833
-2.02059991327962 0.506829997897148
-1.02059991327962 0.502550002932548
-0.0205999132796242 0.500520005822182
0.979400086720376 0.49961000084877
1.97940008672038 0.497219997644424
2.97940008672038 0.496349996328354
3.97940008672038 0.495719993114471
4.97940008672038 0.492930001020432
5.97940008672038 0.493209999799728
6.97940008672038 0.492780005931854
7.97940008672038 0.492509996891022
8.97940008672037 0.491910004615784
9.97940008672037 0.491940003633499
10.9794000867204 0.491110002994537
11.9794000867204 0.491240000724792
12.9794000867204 0.491459995508194
13.9794000867204 0.490909999608994
33 0.490909999608994
};
\addplot [ptcmd, mark=square, dotted, mark options={solid}, opacity=1]
table {%
-36.0205999132796 0.875209999829531
-35.0205999132796 0.871630000323057
-34.0205999132796 0.868489998579025
-33.0205999132796 0.861709998548031
-32.0205999132796 0.858280000090599
-31.0205999132796 0.850920002162457
-30.0205999132796 0.84476999938488
-29.0205999132796 0.835299998521805
-28.0205999132796 0.828350001573563
-27.0205999132796 0.815890000760555
-26.0205999132796 0.80097000002861
-25.0205999132796 0.784329999983311
-24.0205999132796 0.767669996619225
-23.0205999132796 0.742550000548363
-22.0205999132796 0.71807000041008
-21.0205999132796 0.687759998440743
-20.0205999132796 0.657809999585152
-19.0205999132796 0.613550001382828
-18.0205999132796 0.570050004124641
-17.0205999132796 0.526929995417595
-16.0205999132796 0.481029999256134
-15.0205999132796 0.437050002813339
-14.0205999132796 0.391700005531311
-13.0205999132796 0.35436999797821
-12.0205999132796 0.32043000459671
-11.0205999132796 0.29229000210762
-10.0205999132796 0.271319997310638
-9.02059991327963 0.255519998073578
-8.02059991327963 0.24135000705719
-7.02059991327962 0.230179995298386
-6.02059991327962 0.224500000476837
-5.02059991327962 0.217979991436005
-4.02059991327962 0.213680005073547
-3.02059991327962 0.210929995775223
-2.02059991327962 0.207370001077652
-1.02059991327962 0.206790000200272
-0.0205999132796242 0.204989993572235
0.979400086720376 0.204909998178482
1.97940008672038 0.203910005092621
2.97940008672038 0.204869997501373
3.97940008672038 0.203989994525909
4.97940008672038 0.203900003433228
5.97940008672038 0.203159999847412
6.97940008672038 0.203459990024567
7.97940008672038 0.202660006284714
8.97940008672037 0.202530008554459
9.97940008672037 0.20295000076294
10.9794000867204 0.202859997749329
11.9794000867204 0.202820003032684
12.9794000867204 0.202510005235672
13.9794000867204 0.20300999879837
33 0.20300999879837
};
\addplot [ptoamp, empty legend, dashed, mark = None, mark options={solid}]
table {%
0 0
1 2
};
\addplot[ptoamp, mark options={solid}, opacity=1] %
table {%
-36.0205999132796 0.892750024795532
-35.0205999132796 0.889729976654053
-34.0205999132796 0.889789998531342
-33.0205999132796 0.88959002494812
-32.0205999132796 0.887930035591125
-31.0205999132796 0.885179996490479
-30.0205999132796 0.882749974727631
-29.0205999132796 0.878929972648621
-28.0205999132796 0.876439988613129
-27.0205999132796 0.87253999710083
-26.0205999132796 0.870260000228882
-25.0205999132796 0.862489998340607
-24.0205999132796 0.855019986629486
-23.0205999132796 0.847689986228943
-22.0205999132796 0.8367999792099
-21.0205999132796 0.82600998878479
-20.0205999132796 0.808930039405823
-19.0205999132796 0.79026997089386
-18.0205999132796 0.765399992465973
-17.0205999132796 0.738109946250916
-16.0205999132796 0.707340002059937
-15.0205999132796 0.671299934387207
-14.0205999132796 0.632019996643066
-13.0205999132796 0.596530020236969
-12.0205999132796 0.564539968967438
-11.0205999132796 0.540130019187927
-10.0205999132796 0.521589994430542
-9.02059991327963 0.507869958877563
-8.02059991327963 0.50094997882843
-7.02059991327962 0.496679961681366
-6.02059991327962 0.494109988212585
-5.02059991327962 0.492760002613068
-4.02059991327962 0.49115002155304
-3.02059991327962 0.490090012550354
-2.02059991327962 0.490409970283508
-1.02059991327962 0.488900005817413
-0.0205999132796242 0.489299952983856
0.979400086720376 0.488219976425171
1.97940008672038 0.48919004201889
2.97940008672038 0.487670004367828
3.97940008672038 0.487520039081573
4.97940008672038 0.487339973449707
5.97940008672038 0.487769961357117
6.97940008672038 0.487230002880096
7.97940008672038 0.487230062484741
8.97940008672037 0.487249970436096
9.97940008672037 0.487370014190674
10.9794000867204 0.487129986286163
11.9794000867204 0.486929953098297
12.9794000867204 0.486760020256042
13.9794000867204 0.486939966678619
33 0.486939966678619
};
\addplot [ptoamp, mark =triangle, mark size=2, dotted, mark options={solid}, opacity=1]
table {%
-36.0205999132796 0.886299967765808
-35.0205999132796 0.884299993515015
-34.0205999132796 0.882550001144409
-33.0205999132796 0.880819976329803
-32.0205999132796 0.876410007476807
-31.0205999132796 0.874469995498657
-30.0205999132796 0.867940008640289
-29.0205999132796 0.865559995174408
-28.0205999132796 0.860260009765625
-27.0205999132796 0.852190017700195
-26.0205999132796 0.844030022621155
-25.0205999132796 0.836510002613068
-24.0205999132796 0.82166999578476
-23.0205999132796 0.807590007781982
-22.0205999132796 0.791220009326935
-21.0205999132796 0.772459983825684
-20.0205999132796 0.752140045166016
-19.0205999132796 0.722620010375977
-18.0205999132796 0.694820046424866
-17.0205999132796 0.664740025997162
-16.0205999132796 0.630259990692139
-15.0205999132796 0.594059944152832
-14.0205999132796 0.563610017299652
-13.0205999132796 0.531599998474121
-12.0205999132796 0.50543999671936
-11.0205999132796 0.483009994029999
-10.0205999132796 0.464399993419647
-9.02059991327963 0.44925993680954
-8.02059991327963 0.438510000705719
-7.02059991327962 0.430209994316101
-6.02059991327962 0.422490000724792
-5.02059991327962 0.418109953403473
-4.02059991327962 0.414849996566772
-3.02059991327962 0.411710023880005
-2.02059991327962 0.409309983253479
-1.02059991327962 0.407339990139008
-0.0205999132796242 0.405369997024536
0.979400086720376 0.40421998500824
1.97940008672038 0.404070019721985
2.97940008672038 0.403169989585876
3.97940008672038 0.402509987354279
4.97940008672038 0.401800036430359
5.97940008672038 0.402419984340668
6.97940008672038 0.401669979095459
7.97940008672038 0.401910006999969
8.97940008672037 0.401660025119781
9.97940008672037 0.401130020618439
10.9794000867204 0.401430010795593
11.9794000867204 0.400780022144318
12.9794000867204 0.401399970054626
13.9794000867204 0.40091997385025
33 0.40091997385025
};
\end{axis}
\end{tikzpicture}